\documentclass[letterpaper,twocolumn,10pt]{article}
\usepackage{usenix}

\usepackage{tikz}
\usepackage{amsmath}
\usepackage[utf8]{inputenc} 
\usepackage[T1]{fontenc}    
\usepackage{hyperref}       
\usepackage{url}            
\usepackage{booktabs}       
\usepackage{amsfonts}       
\usepackage{nicefrac}       
\usepackage{algorithm}
\usepackage{microtype}      
\usepackage{xcolor}         
\usepackage{graphicx}
\usepackage{algpseudocode}
\usepackage{xcolor}
\usepackage{enumitem}
\usepackage{caption}
\usepackage{subcaption}
\usepackage[normalem]{ulem}
\usepackage{placeins}
\usepackage{enumitem}
\newcommand{\YF}[1]{\textcolor{blue}{#1}}

\newcommand{\red}[1]{#1}

\newcommand{\rv}[1]{#1}

\begin{document}

\date{}

\title{PlanetServe: A Decentralized, Scalable, and Privacy-Preserving Overlay for Democratizing Large Language Model Serving}

\author{
\textnormal{Fei Fang}$^{\dagger,}$\thanks{Co-primary authors.},\;
\textnormal{Yifan Hua}$^{\dagger,}$\footnotemark[1],\;
\textnormal{Shengze Wang}$^{\dagger,}$\footnotemark[1],\;
\textnormal{Ruilin Zhou}$^{\dagger}$,\;
\textnormal{Yi Liu}$^{\dagger}$,\;
\textnormal{Chen Qian}$^{\dagger}$,\;
\textnormal{Xiaoxue Zhang}$^{\ddagger}$ \\
\textit{$^{\dagger}$University of California, Santa Cruz \quad
$^{\ddagger}$University of Nevada, Reno}
}
\maketitle

\begin{abstract}
While significant progress has been made in research and development on open-source and cost-efficient large language models (LLMs), serving scalability remains a critical challenge, particularly for small organizations and individuals seeking to deploy and test their LLM innovations. Inspired by peer-to-peer networks that leverage decentralized overlay nodes to increase throughput and availability, we propose PlanetServe, an LLM serving overlay that harnesses computing resources from decentralized contributors. We identify four key research problems inherent to enabling such a decentralized infrastructure: 1) overlay network organization; 2) LLM communication privacy; 3) overlay forwarding for resource efficiency; and 4) verification of serving quality. This work presents the first systematic study of these fundamental problems in the context of decentralized LLM serving. Evaluation results from a prototype implemented on a set of decentralized nodes demonstrate that PlanetServe achieves a latency reduction of over 50\% compared to the baseline design without overlay forwarding. Furthermore, the security features introduce minimal overhead to serving latency and throughput. We believe this work pioneers a new direction for democratizing and scaling future AI serving capabilities. 
\end{abstract}

\vspace{-2ex}
\section{Introduction}
\vspace{-1ex}
The rapid development and proliferation of open-source, cost-efficient large-language models (LLMs), such as Google's Gemma \cite{gemma}, DeepSeek-R1 \cite{deepseek}, and Meta's LLaMA series \cite{touvron2023llama}, have significantly lowered the barrier to AI deployment. This progress makes it feasible to run advanced language models on personal workstations and local servers with consumer GPUs. Despite this accessibility, \textbf{serving scalability remains a critical challenge for small organizations and individuals to deploy LLM innovations}. Individual researchers, university laboratories, and start-up companies often find it prohibitively expensive and technically challenging to deploy models at scale, particularly when popularity surges. For instance, some promising models initially provided excellent performance at low cost, but later suffered severe performance degradation when user demand suddenly increased \cite{DeepseekSlow}. 
Traditional cloud-based infrastructures are not always affordable or accessible for smaller organizations and individuals. 

We propose PlanetServe, a decentralized overlay infrastructure designed for large-scale LLM deployment. Inspired by PlanetLab \cite{planetlab}, a well-known peer-to-peer (P2P) network allowing researchers to deploy and test Internet applications across distributed nodes, PlanetServe harnesses computing resources from a diverse range of contributors, including individuals, academic institutions, and small organizations. Participants in PlanetServe volunteer computing power and network resources, and in return gain the ability to deploy LLMs and AI-driven services on this shared, decentralized network.

PlanetServe provides the following main advantages:
\begin{enumerate}
\vspace{-1.5ex}
\item 
\textbf{Scalability, resiliency, and cost-efficiency.} 
PlanetServe can dynamically adapt to rapid fluctuations in demand, eliminating the performance bottlenecks commonly associated with centralized cloud deployments that depend on rigid, pre-defined service agreements. Idle computing resources 
can be utilized instead of expensive dedicated cloud instances. \red{PlanetServe increases scalability on both computation and networks. Hence, it can resolve the scalability problem for computation (e.g., large models), networks (e.g., small models), or a mixed situation.}

\vspace{-1.6ex}
\item \textbf{Democratization of AI innovation development.}  PlanetServe significantly lowers the barrier to independent researchers, developers, and academic institutions being able to deploy and experiment with their LLM innovations at scale. Hence, it encourages innovation from a wider and more diverse contributor base.
\vspace{-1.5ex}
\item \textbf{Enhance user privacy.} Conventional cloud deployments often centralize user prompts and associated private information, creating a single point of trust and potential vulnerability. PlanetServe addresses this by incorporating anonymous overlay networks, preventing model-serving nodes from easily linking user prompts to individual identities, thereby enhancing user privacy.
\vspace{-2.5ex}
\item PlanetServe naturally inherits benefits that are common to decentralized overlay systems, such as service redundancy, geographic distribution of resources, and load balancing among heterogeneous servers. 
\vspace{-1ex}
\end{enumerate}
\vspace{-1ex}

PlanetServe represents a novel approach to deploying LLMs and other advanced AI services, democratizing large-scale AI capabilities while enhancing user privacy. Although model serving today is largely concentrated in large-scale cloud deployments, the software ecosystem~\cite{mlc-llm} has demonstrated that inference can be effectively deployed across diverse devices. This trend suggests a future inference landscape should also be determined by cost efficiency, privacy, and availability. A decentralized infrastructure like PlanetServe is well-suited to this evolution, enabling participants to contribute heterogeneous resources align with their needs (e.g., low-latency private inference or specialized model serving).

In designing and implementing PlanetServe, this work addresses four core research challenges that are critical for any decentralized LLM serving infrastructure: 1) Overlay network organization. 2) Anonymous communication to protect user privacy. 3)  Overlay forwarding for serving efficiency. 4) Decentralized verification of compliant model serving.
\textbf{To our knowledge, this is the first work to systematically study these core challenges for achieving LLM serving scalability via a decentralized overlay network independent of centralized cloud control.} 

We implemented  PlanetServe's core functions as a prototype consisting of decentralized nodes and tested the prototype in a public cloud without central control. This approach provided a controllable testbed, necessary as we do not have direct access to a large set of physical decentralized servers across the Internet. We let each node add synthetic latency to every packet for the wide-area Internet conditions. 
Upon acceptance of this paper, we will make the prototype implementation and datasets open-sourced. 


This work focuses on decentralized LLM serving over an overlay network and assumes model training is completed by developers. We exclude decentralized training~\cite{ye2024openfedllm}, collaborative serving of a LLM instance~\cite{borzunov2023petal,DistServe2024,SplitWise2024}, and cloud-based distributed serving with centralized scheduling~\cite{sglang24,Preble25,DistServe2024,SplitWise2024,CacheGen2024}. 

The rest of this paper is organized as follows. Sec.~\ref{sec:model} presents the system model and design objectives of PlanetServe. Sec.~\ref{sec:design} details the system design. Sec.~\ref{sec:security} and Sec.~\ref{sec:evaluation} provide the security analysis and performance evaluation, respectively. Sec.~\ref{sec:discuss} discusses design insights and practical considerations, followed by Sec.~\ref{sec:limitations}, which outlines the limitations of the current approach. Sec.~\ref{sec:related} reviews related work. Finally, Sec.~\ref{sec:conclusion} concludes the paper and discusses future directions.

\vspace{-2.5ex}
\section{System Models and Objectives}
\label{sec:model}
\vspace{-2ex}

\begin{figure}[t]
    \centering
    \includegraphics[scale=0.45]{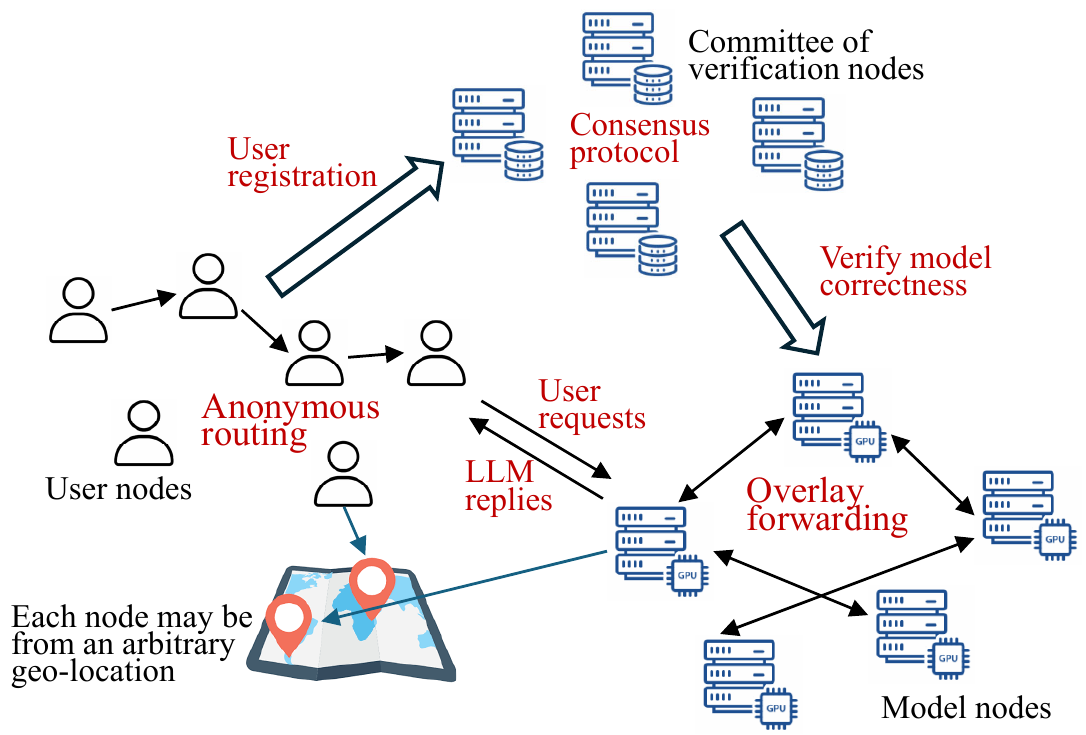}
    \vspace{-2ex}
    \caption{Overview of PlanetServe.} 
    \vspace{-3ex}
    \label{fig:overview}
\end{figure}


\subsection{Network Model}
\vspace{-1ex}
PlanetServe operates as an overlay network composed of end hosts, each plays as one or more of the following roles:
\begin{enumerate}[left=0pt]
\vspace{-2ex}
    \item \textbf{User node} represents end-users (clients) seeking access to LLM services while preserving privacy. 
    \vspace{-2ex}
    \item \textbf{Model node} executes LLM inference tasks to serve user requests. These nodes typically run on servers with GPU resources, contributed by participating organizations or individuals. 
    Each model node serves one or more complete LLM instances. \red{Collaborative inference across multiple nodes for a single request \cite{borzunov2023petal,DistServe2024} is not considered.}
    \vspace{-2ex}
    \item \textbf{Verification node.} To maintain correct LLM serving and defend against security attacks, PlanetServe includes a committee of decentralized verification nodes, far fewer than users and model nodes. Each verification node runs on a server owned by an organization or individual with certain credentials, such as a well-recognized university research team. 
 They are to ensure that each model node behaves legitimately 
    and manage user and model node information. 
    This committee runs a BFT consensus protocol \cite{TendermintBFT} to ensure information correctness and consistency. 
    \vspace{-4ex}
\end{enumerate}

A single physical device may operate in two or all roles in PlanetServe. All communications between nodes in PlanetServe are via TCP, secured with TLS. 
User nodes may join or leave the network at any time and hence network churn can happen. Model and verification nodes are more stable but they still can fail or be compromised by attackers. 

\vspace{-2ex}
\subsection{\red{Cost Advantage and Incentive Model}}
\label{sec:incentive}
\vspace{-1ex}

\red{\textbf{Cost advantage of PlanetServe.} The cost advantage of PlanetServe compared to commercial public cloud is mainly from two facts. First, computation resources remain dramatically underutilized in organizations. Hence, the cost of contributing servers to PlanetServe is very low. Many organizations deployed GPU servers for ML training, but the servers sit idle outside of training times. It has been reported that the average server utilization rate is between 12\%-18\% of capacity \cite{underulti}. 10 million servers are estimated to sit completely idle, representing \$30 billion in wasted capital \cite{underulti}. These idle servers can be contributed to PlanetServe and form a large-scale serving infrastructure. Second, using community-contributed GPUs for inference can dramatically cut costs compared to commercial cloud platforms. Measurements show that the energy cost of running LLM inference on GPUs such as the V100 and A100 is relatively modest~\cite{samsi2023words},  implying that the high hourly rates of public cloud instances are driven largely by infrastructure and pricing overheads rather than intrinsic hardware or energy expenses. Recent work on scalable inference similarly reports $20\times$ lower cost when serving LLMs on consumer GPUs versus AWS instances \cite{ScaleLLM}. In addition, many consumer GPUs or GPUs that have been given away from training can still serve moderately sized LLMs (7B–13B parameters).} \rv{Notably, this comparison focuses on cost efficiency, whereas commercial public clouds provide additional services beyond cost.}

\red{In summary, prior measurements and cost analyses suggest that LLM serving and testing using private hardware can be an order of magnitude cheaper than public cloud if the hardware is fully utilized. However, scaling private hardware is challenging, and hardware utilization is very low. PlanetServe addresses both the scalability and under-utilization problem of private hardware by allowing different organizations to contribute their resources and share them with each other.} 

\red{PlanetServe uses a \textbf{reputation-based incentive model} to encourage participation and cooperation. The incentive model only applies to the organizations that contribute model nodes to the system. There is no system-wide incentive mechanism for users. However, model deployers can use their own policies to regulate users who can access their LLMs.}  

\red{\textbf{Incentive for participation.} Each organization participates in the system by contributing a number of model nodes. Model nodes from the same organization share the same reputation score $\lambda$, which will be updated during the system execution, as detailed in Sec.~\ref{sec:veri}. If the reputation score is above a threshold, the organizer is allowed to deploy their own LLM to the system. To measure the resources it is allowed to use for deploying its LLM, each organization also has a contribution credit, which reflects the resource time it contributes to the system. The contribution credit is calculated proportionally to the cost of renting servers from a public cloud. For example, if an organization has contributed 5 servers that have been serving for 30 days in PlanetServe, it can deploy its LLM to PlanetServe that runs on 30 servers with similar computing resources for 5 days. All reputation scores and contribution credits are assigned and updated by the committee of verification nodes running a consensus protocol.}

\red{\textbf{Incentive for cooperation.} If model nodes do not follow the PlanetServe protocols in model serving and request forwarding, their reputation scores will be reduced, which will hurt their eligibility to use other resources in PlanetServe. The verification node committee is responsible for ensuring that the model nodes correctly run the models they promise to serve and for maintaining the reputation scores of the model nodes, as detailed in Sec.~\ref{sec:veri}.}

\vspace{-3ex}
\subsection{Threat Model and Assumptions}
\vspace{-1.5ex}
Consistent to many existing works \cite{IS-NSDI07,LAP12,garlic}, this work focuses on \textbf{practical low-delay security and anonymity}, rather than perfect security at all costs. We assume the following attacks by different nodes in PlanetServe. 

\begin{itemize}[leftmargin=0pt,itemsep=0.2ex,parsep=0pt,topsep=0.2ex,partopsep=0pt]
    \item Malicious user nodes may drop packets they are responsible for forwarding. They may try to deanonymize the sender of a request and inspect packet contents. We also assume user nodes that can collude to conduct these attacks.   
    \item A dishonest model node may drop user requests on purpose or reply with incorrect model results to mislead users. This includes manipulating outputs or substituting the agreed-upon model with a cheaper, lower-quality one. For example, the node claims to serve an 8B-parameter LLM but runs a 1.5B model instead to conserve resources. It will try to differentiate normal requests from users and verification requests to treat them differently. Even though a model node does not perform active attacks, it might attempt to link the received prompts and their user nodes. It might falsely claim resource exhaustion to refuse service.  \red{A malicious model node may also be curious about sensitive prompt content, such as users' personal data or proprietary information.}
    \item A compromised verification node could provide wrong information to users or false verification results for model nodes. 
    We assume an adversary, potentially leveraging Sybil attacks to control multiple verification nodes.
   No more than $f$ out of $N=3f+1$ verification nodes can be compromised following the Byzantine fault tolerance (BFT) assumption.
   This threshold is based on the fact that verification nodes are run by decentralized entities, making the cost of compromising a sufficient fraction extremely high. 
\end{itemize}

\vspace{-3.5ex}
\subsection{Design Objectives}
\vspace{-1.5ex}

PlanetServe is designed to achieve the following objectives:

\textbf{Performance:} The key performance goals include providing \textbf{low-latency LLM services} with high scalability. In addition, PlanetServe aims to achieve \textbf{load balancing}, preventing individual nodes from being overwhelmed. 

\textbf{Security, privacy, and availability:}
PlanetServe aims to preserve user anonymity and protect the confidentiality of prompt content and model responses. 
It is designed to prevent model nodes or other users from linking user identities (e.g., IP addresses \red{or public keys}) to their requests.
\red{For privacy-sensitive workloads, PlanetServe leverages NVIDIA Confidential Computing on Hopper and Blackwell GPUs to run inference inside trusted execution environments (TEEs), ensuring that model nodes cannot access prompts or proprietary information.~\cite{NVH100, nvidia-secure-ai-blackwell-hopper}. In addition, PlanetServe should provide high service availability under node churn and attacks and ensure model node provide accurate responses.}

\vspace{-3.5ex}
\section{System Design of PlanetServe}
\vspace{-2ex}
\label{sec:design}


\subsection{Overview}
\vspace{-1ex}
The overview of PlanetServe is illustrated in Fig.~\ref{fig:overview}. 

The user nodes form a dynamic overlay network. Upon joining PlanetServe, a user node registers its \red{current IP address and} public key with the verification nodes, whose IP addresses and public keys are public information. \red{The public key serves as the identifier.} It then downloads a list of active user and model nodes from a verification node, and is signed by more than 2/3 of the verification nodes \cite{PBFT99}. 
\red{Verification nodes may choose to divide the whole system into multiple regions and create a list of users and model nodes for each region, only when the number of users in each region is sufficiently large to hide the requester's identity, for example, $>1000$ users \cite{Intruder,hordes00}.} 
PlanetServe allows a user to use other users as relays for \textit{anonymous communication}: each user request is forwarded along an overlay path to hide the address of the original sender, and the LLM reply will also be forwarded along the anonymous path back to the requesting user. 

Each model node runs an instance of an LLM, which is deployed as a portable isolated environment, such as a container \cite{Container2017}, provided by the LLM developer. 
One or more LLMs are deployed in the network, and each user request specifies which LLM it is requesting. 
A user request can be responded to by any model node that serves the same LLM as its request target. 
Model nodes use \textit{overlay forwarding} to achieve load balancing and KV cache reuse among them. 

The committee of verification nodes uses a recent BFT consensus protocol \cite{TendermintBFT} to ensure consistency.
The committee maintains a list of user nodes registered to PlanetServe, including their public keys, and a list of model nodes in PlanetServe, including their public keys and reputation scores. The reputation score of a model node quantifies the quality of its LLM responses and will be defined in Sec.~\ref{sec:veri}. \red{Verification nodes periodically request the model nodes anonymously
and update their scores based on the responses.} 


\vspace{-2ex}
\subsection{Anonymous Overlay \red{and Content Privacy}}
\label{sec:anonymous}
\vspace{-1ex}
The anonymous overlay of users in PlanetServe allows each user node to request the LLM service without revealing its IP address. We summarize the requirements of the anonymous overlay as follows: \red{1) User node anonymity: other nodes cannot link a prompt with the requesting user identity. 2) Confidentiality of prompts and responses: the contents can only be visible to the requester and model node. 3) Low overhead on user nodes. In particular, we do not want to use public-key cryptography for prompts and responses, which introduces large overhead at relay nodes. 4) A single anonymous path is susceptible to relay failures under network churn. PlanetServe uses multiple paths to provide reliable message delivery.}

We analyze typical anonymous overlay communication protocols \cite{Tor04,Han2006RR,IS-NSDI07,LAP12,garlic}, including Onion routing and sliced routing \cite{Han2006RR,IS-NSDI07,garlic}, in the Appendix and find that none of them fit the above requirements. \red{If we apply Onion to the overlay network in PlanetServe, there are a few main challenges. 1) The first layer of Onion nodes (called guards in Tor) will definitely know the sending user of a request, which should be avoided in PlanetServe. 2) The failure probability of an Onion path increases exponentially with the path length, especially for an overlay network of users with a high churn rate. 3) Each relay needs to perform public-key decryption on the entire message, introducing non-trivial overhead for users.}

We propose a novel approach that achieves the advantages of both Onion and sliced routing and fits the requirements of PlanetServe exactly. In PlanetServe, each user uses Onion routing to establish $n$ proxies. In this process, path failures and redundancy do not cause high resource waste because the establishment message is very short. After proxies are established, the user and model nodes rely on sliced routing for prompt and response messages.

\textbf{Using S-IDA.} 
We briefly present an existing protocol Secure Information Dispersal Algorithm (S-IDA) \cite{SIDA93} that is essential to achieve message confidentiality, anonymous routing, and failure resilience in PlanetServe. 
Secure IDA (S-IDA) \cite{SIDA93,garlic} 
improves the security level of classic IDA \cite{Rabin89} by combining it 
with symmetric
encryption.
The sender applies the following steps. 1) Encrypts $M$ by an AES key $K$, getting $\{M\}_K$.
2) Splits $\{M\}_K$ into $n$ fragments, $M_1, M_2, ...,
M_n$ by a $k$-threshold Rabin's IDA.
3) Splits $K$ into $n$ fragments, $K_1, ..., K_n$ by a $k$-threshold Shamir's secret sharing (SSS) \cite{Shamir1979}.
4) Constructs $n$ messages called \emph{cloves}, $C_1, C_2, ..., C_n$,
where $C_i$ includes two fields $M_i$ and $K_i$.
5) Sends the cloves to the receiver using $n$ different paths. 
By receiving at least $k$ cloves, the receiver:
1)  Recovers $\{M\}_K$ using Rabin's IDA and the key $K$ by SSS.
2) Decrypts $\{M\}_K$ and gets $M$.
Only when a node collects $k$ cloves, it can
recover the key and decrypted $\{M\}_K$. Without the complete key, it can
only see the ciphertext related to $\{M\}_K$.
We call the above process $(n,k)$ S-IDA. 

\begin{figure}[t]
    \centering
    \includegraphics[scale=0.46]{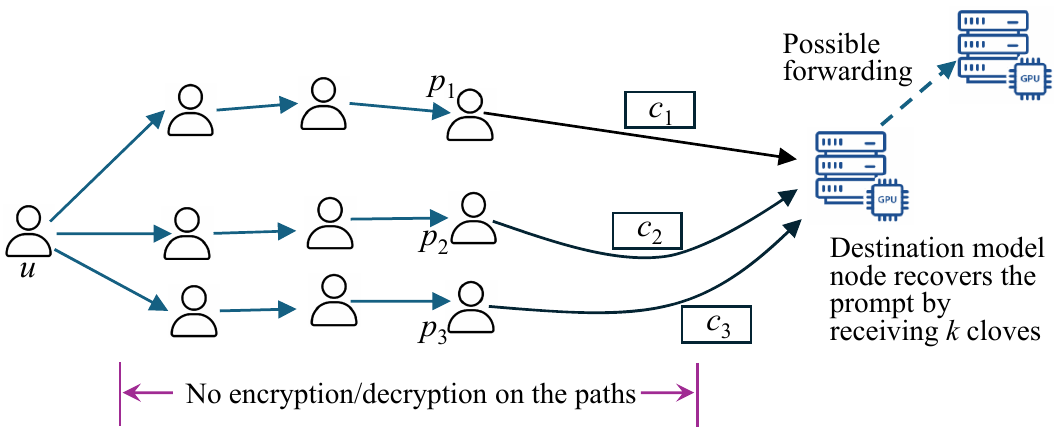}
    \vspace{-2ex}
    \caption{Process of querying the model node.} 
    \vspace{-3ex}
    \label{fig:query}
\end{figure}

\begin{figure}[t]
    \centering
    \includegraphics[scale=0.43]{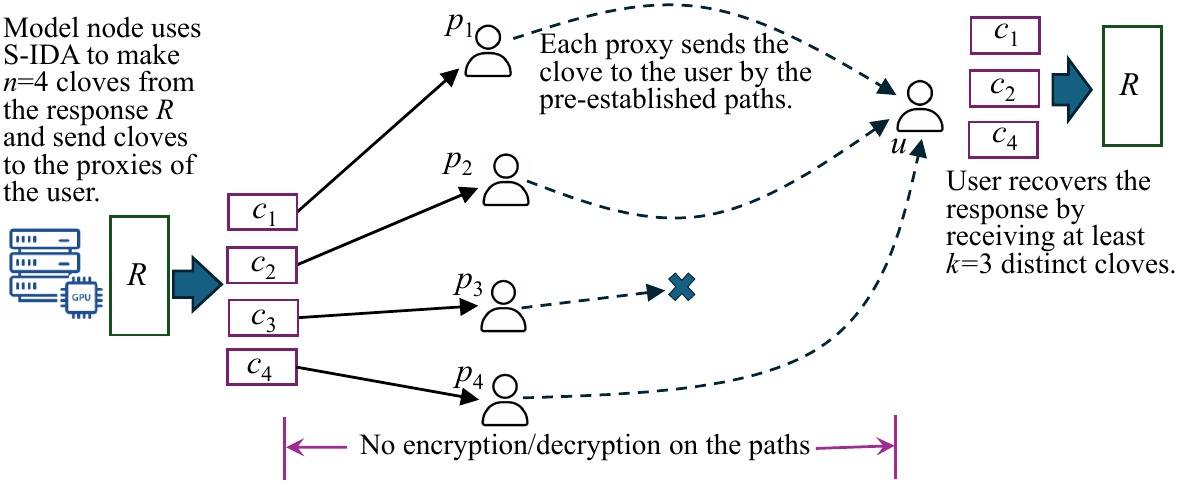}
    \vspace{-4ex}
    \caption{Process of sending the reply message from the model node to the user.} 
    \vspace{-4ex}
    \label{fig:reply}
\end{figure}
We present the detailed anonymous protocol. 

\textbf{Step 1. Preparation.} 
A new user $u$ contacts an arbitrary verification node to download a list of overlay users, called the \textit{user list}, and a list of model nodes, called the \textit{model node list}, which are signed by more than 2/3 verification nodes. 
Each entry in the list includes the public key and IP address. 
PlanetServe determines $n$ and $k$ ($n>k$) for $(n,k)$ S-IDA. 

\textbf{2. Proxy and path establishment.} Each user node, say $u$, needs to find $N$  proxies ($N \geq n$). 
To find each proxy, the user $u$ selects $l$ other users from the user list, where $l$ is the path length. 
\red{Conventional wisdom indicates that anonymous paths with three routers in Tor 
can achieve an appropriate balance between security and latency \cite{PathTor10}. Hence PlanetServe uses $l=3$.} 
The user $u$ then builds an Onion path using the public keys of these 3 selected users and uses the hash value of both $u$ and the last user on the path to be the path session ID $I$.
The last user $p$ on the path becomes a proxy of $u$, and every node on the path stores the predecessor and successor together with the path session ID $I$. 
The above process might fail due to user dynamics but $u$ can easily try other paths.
By repeating the above process multiple times, each user $u$ can establish $N$ proxies using Onion paths.  In later LLM prompts and responses, no public-key cryptographic operations are needed on the paths, because the relay users will use the stored path ID, predecessors, and successors for forwarding. 

\textbf{3. Prompt messages.} When a user node $u$ plans to send a prompt or query message $Q$ to PlanetServe, it first selects a model node from the model node list and writes its IP as the destination. As illustrated in Fig.~\ref{fig:query}, $u$ prepares $n$ slices (called \textit{cloves}) of $Q$ using $(n,k)$ S-IDA and sends them to $n$ proxies respectively using the established paths.
The query message $Q$ includes only the prompt and model node IP without any information about $u$. 
The application-layer header of each clove includes the path ID $I$ for the corresponding proxy but does not include the proxy IP. 
When each proxy receives the clove, it directly sends the clove to the destination model node, which is not anonymous. The prompt message also includes the IP addresses of at least  $n$ proxies of $u$, which are revealed to the model nodes when it receives $\geq k$ cloves.

\textbf{4. Response messages.} 
As shown in Fig.~\ref{fig:reply}, each response message is made into $n$ cloves using $(n,k)$ S-IDA. Then the model node sends $n$ cloves to $n$ proxies of $u$ respectively. Each proxy then forwards the received clove to $u$ using the pre-established anonymous path. 
When $u$ receives more than $k$ cloves, it can decrypt the reply message using S-IDA. The content of the response is confidential unless an attacker controls users that appear on $k$ or more of these $n$ paths. Even in that case, the attacker needs to use brute force to decode S-IDA because these $n$ paths have different IDs. 

\red{\textbf{Content privacy.} PlanetServe supports model nodes with Confidential Computing (CC) hardware, including NVIDIA Hopper and Blackwell GPUs such as H100 \cite{NVH100,nvidia-secure-ai-blackwell-hopper}.  This will allow users to choose between two privacy tiers: a default user-anonymous mode as stated above, and an optional content-privacy mode, in which both the identity and the prompt content remain protected from the node operator. For sensitive scenarios such as resume editing, healthcare, or finance, users can opt for content anonymity. Achieving content privacy protection requires hardware-based Trusted Execution Environments (TEEs) because software-only encryption can protect data in transit but not in use—once plaintext resides in GPU memory for inference, it is otherwise accessible to the host OS or hypervisor. The H100 CC mode addresses this by booting in a verified state with an on-die hardware root of trust, ensuring only NVIDIA-signed firmware is loaded, and by running inference inside an isolated GPU TEE in which all PCIe/NVLink traffic is encrypted with AES-GCM-256 and rolling IVs to prevent replay attacks. Data is exchanged with the CPU TEE (e.g., AMD SEV-SNP or Intel TDX) only through encrypted bounce buffers. Hence, plaintext is never exposed outside the enclave. Before processing, the GPU can be remotely attested via NVIDIA’s attestation service to verify its identity, firmware integrity, and CC configuration. 
Each LLM is launched in a Confidential VM (CVM) with the signatures of the verification committee, using the CC mode. User nodes establish end-to-end TLS sessions directly with the CVM, ensuring that neither the host nor the hypervisor can observe the prompt content. 
We show the performance results via implementation in Sec.~\ref{sec:CC-latency}. 
}
\vspace{-2.5ex}
\subsection{Overlay Forwarding among Model Nodes}
\label{sec:overlay}
\vspace{-1ex}

\begin{figure}[t]
    \centering
    \includegraphics[width=0.99\linewidth]{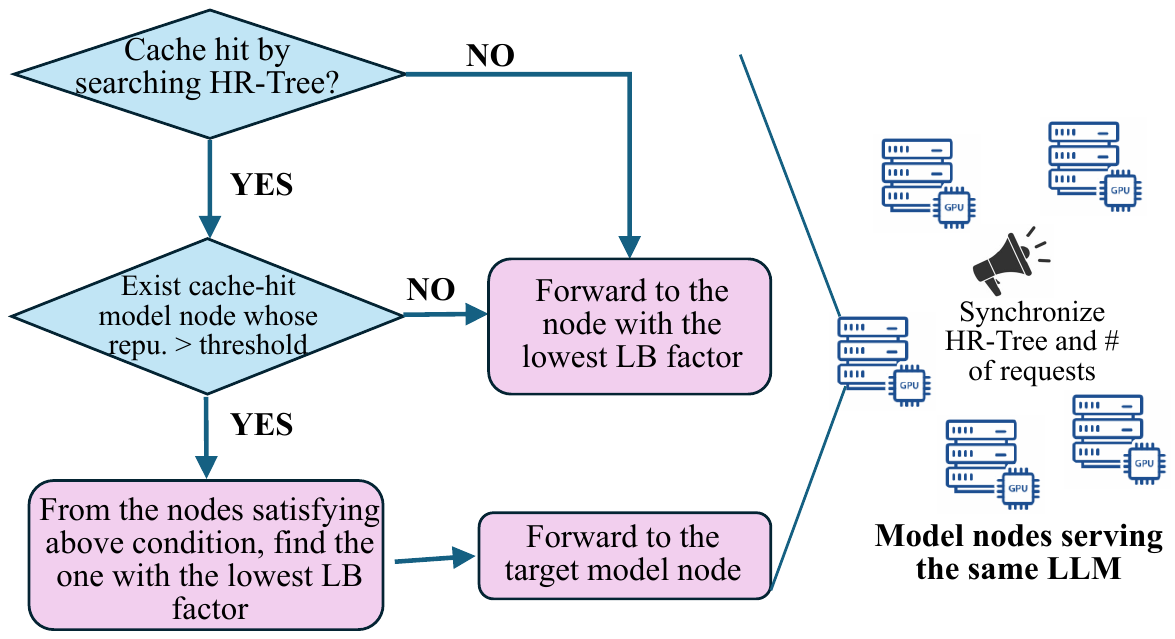}
    \vspace{-2ex}
    \caption{\red{The logic of overlay forwarding}}
    \vspace{-3ex}
    \label{fig:forward}
\end{figure}

All model nodes serving the same LLM form a logical group. 
Upon receiving a request from a user, a model node can serve this request or forward the request to another node. 
The forwarding logic, illustrated in Fig.~\ref{fig:forward}, is designed to consider both KV cache reusing and load balancing objectives. The detailed design includes the following components.

\textbf{Hash-Radix tree.}  
Recent studies \cite{sglang24,Preble25,cheng2025distributedllm} show that sharing and reusing KV cache among different GPUs can significantly reduce the average latency for LLM requests. However, prior work such as SGLang \cite{sglang24} and Preble \cite{Preble25} assumes a cloud environment that has a centralized scheduler or request router. This central entity receives all requests and distributes them to the appropriate GPUs. PlanetServe lacks such a central coordinator. The centralized scheduler \cite{sglang24,Preble25} uses a radix tree \cite{Morrison1968} for searching existing KV caches that can be reused for the current prompt, based on matching prefixes. Each tree node in a radix tree represents a prefix. 

In a decentralized system like PlanetServe, naively requiring every model node to maintain a full copy of the radix tree built from the KV cache information of all model nodes in the group introduces significant scalability problems: 1) High memory overhead: each node needs to spend substantial memory space to store the aggregated radix tree, including the metadata for the KV cache from all other nodes; and 2) high communication overhead: each node needs to broadcast updates about the prefixes of its local KV cache to all other nodes in the group. To resolve these problems, we propose a new distributed data structure called a Hash-Radix tree (HR-tree), detailed below.



\begin{figure}[t]
    \centering
    \includegraphics[width=0.9\linewidth]{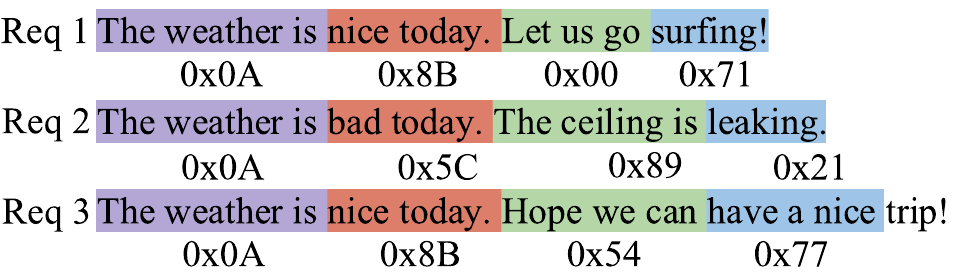}
    \vspace{-2ex}
    \caption{Pre-processing for an HR-tree}
    \vspace{-3ex}
    \label{fig:hrt1}
\end{figure}

\begin{figure}[t]
    \centering
    \includegraphics[width=0.9\linewidth]{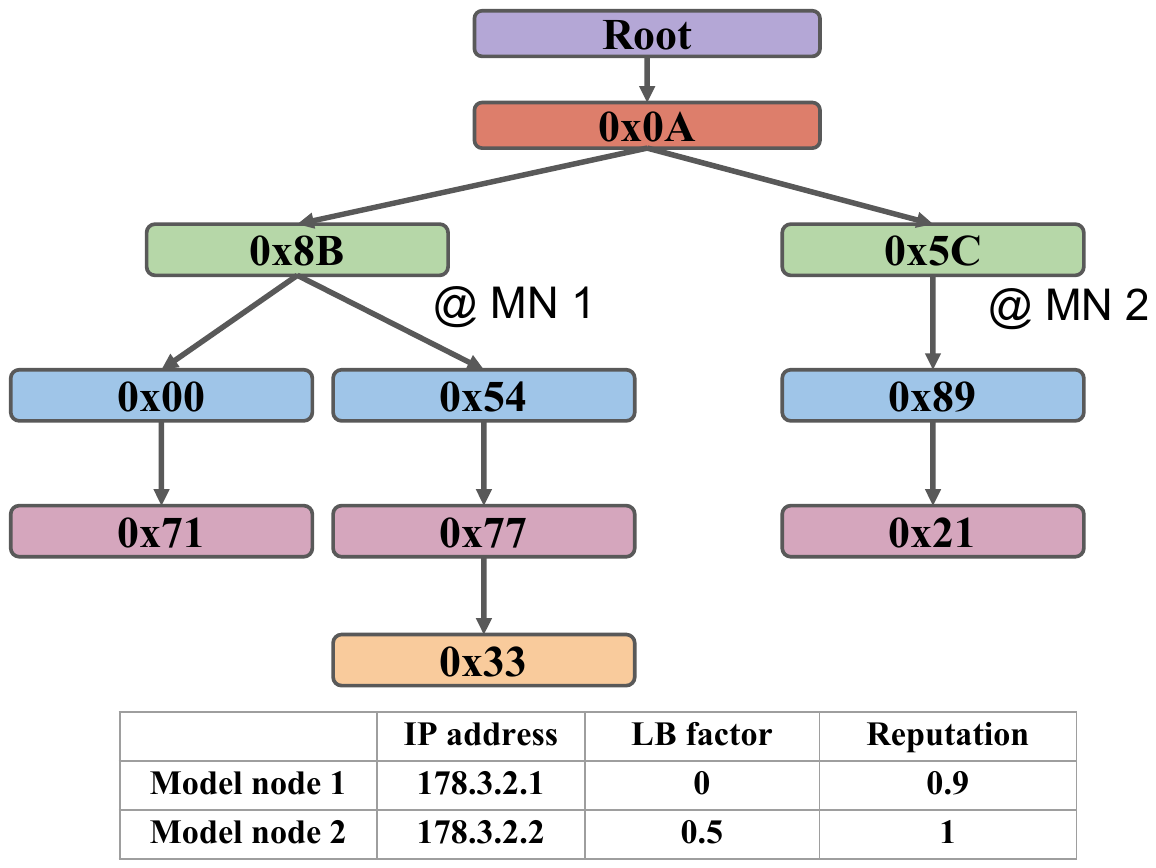}
    \vspace{-1ex}
    \caption{\red{The structure of a Hash-Radix tree}}
    \vspace{-3ex}
    \label{fig:hrt2}
\end{figure}



The design of the HR-tree is motivated by cuckoo filters \cite{cuckoofilter2014}, which replace hash table entries with fingerprints to reduce memory cost. 
An HR-tree represents the aggregated  KV cache state from \textit{all overlay model nodes in the group}. A request prompt can be efficiently searched in the HR-tree. 
If the prompt matches a \textit{sufficiently long} prefix path in the HR-tree, the HR-tree returns the model node that holds the KV cache corresponding to the matched prefix. Otherwise, the HR-tree reports no match. 

The HR-tree construction involves a pre-processing step, shown in Fig.~\ref{fig:hrt1}, where each prompt is divided into variable-length chunks.
Each chunk is converted into a hash value computed by a universal hash function H. The example uses 8-bit hash values such as 0x0A and 0x8B. The length of each chunk $i$ in the number of tokens, $l_i$, is stored in an array $L$ that is computed by a deterministic algorithm called Sentry,  detailed in the Appendix.
The HR-tree is then built based on the sequence of hash values, as shown in Fig.~\ref{fig:hrt2}. Each tree node stores the hash value of a chunk. Two sibling nodes indicate that their corresponding prompts share the same prefix up to the parent node but differ thereafter. 

When each model node receives a user request that includes a prompt, it searches the prompt in the HR-tree. The search algorithm first applies pre-processing to divide the query prompt into chunks and compute their hashes.
The algorithm then traverses a tree structure, starting from the root node. For each chunk of the querying prompt, the algorithm attempts to find a child node with the same hash value as the chunk. 
When there is no child node matching the hash value of the next chunk, the search stops. The HR-tree returns the search depth $d$. If $d$ exceeds a threshold, we consider it a match. Each tree node also includes a set of pointers, each referencing an entry of a separate table. 
\red{The entry represents a model node that holds the corresponding KV cache of the prefix. Each entry in the table includes the node's IP, load balance factor, and reputation score.}

Similar to Bloom and cuckoo filters \cite{Bloom1970,cuckoofilter2014}, an HR-tree might also introduce \textit{false positives}: a prompt that does not  match a prefix can be judged as a match by the HR-tree. For a hash value with 8 bits, the probability of a hash collision is 1/256. However, a false positive 
requires matching $d$ tree nodes, so the false positive rate is as small as $1/256^d$.

\red{\textbf{State synchronization.} 
For each model node in a group, it periodically broadcasts 
the local updates of its HR-tree,
%
each node keeps a snapshot of its HR-tree and the following updates after the snapshot. 
The node periodically sends a minimal but necessary update to all nodes in the group.} \rv{Temporary inconsistencies (e.g., due to churn) during synchronization may reduce cache hit rates without affecting correctness since routing is constrained to nodes running the requested model.}


\textbf{Load balancing} is another design consideration alongside KV cache reuse. Its main objectives are to prevent overload and  ensure each model serves enough requests to establish a reputation score. 
As shown in Fig.\ref{fig:forward}, 
target nodes are selected by minimizing the load-balance (LB) factor, 
$F_{LB} = L * (Q / C)$, 
where $L$ is the moving average of service latency, $Q$ the queued request number, and $C$ is the capacity of the number of concurrent requests. The moving average follows RTT estimation with $\alpha=1/8$. LB factors are computed locally, audited by the verification committee, and periodically broadcast to other nodes. \rv{Consequently, request routing based on $F_{LB}$ adapts to runtime variations across model nodes. As a node's hardware performance degrades or if a node is inherently slower (e.g., consumer-grade GPUs), its $L$ value increases, which redirects incoming requests to nodes with higher effective capacity.}



\textbf{Forwarding logic.} The overlay forwarding logic, executed by \textbf{every model node} upon receiving a user request, determines how to forward the request, as illustrated in Fig.~\ref{fig:forward}. The model node first searches the request's prompt in its HR-tree. This search yields either a match or a miss.  
If the search result is a miss,  the node prioritizes load balancing 
and forwards the request to the model node with the \red{lowest LB factor}. 
If the search result is a match, the HR-tree identifies a list of model nodes that hold reusable KV caches. 
If there is one or more model nodes left, we pick the one with the \red{lowest LB factor} and forward the current request to it. 

\textbf{Session affinity for consecutive prompts.} 
Every model node includes its  IP address in the response message. For subsequent prompts from the same user session, the user routes these prompts specifically to that designated model node, routing them via the anonymous overlay. This policy prioritizes KV cache reuse for consecutive prompts within a session.



\vspace{-2ex}
\subsection{Model verification}
\label{sec:veri}
\vspace{-1ex}
The committee of verification nodes is responsible for ensuring that the model nodes correctly run the models they promise to serve by maintaining their reputation scores. 
The verification nodes run a BFT-like consensus protocol \cite{TendermintBFT} to ensure data consistency and authenticity in the committee. 
The fundamental verification check is to compare a response from a target model node for a specific prompt against a reference result generated locally using the same LLM. However, verifying every user response is impractical due to scalability  and potential user privacy issues.

PlanetServe introduces a sampling-based verification strategy. Instead of monitoring all responses, verification nodes periodically send challenge prompts to model nodes and verify their responses. 
These challenge prompts are routed by the anonymous overlay; hence, \textbf{model nodes cannot distinguish verification probes from regular user requests}.
Verification nodes then assign each model node a reputation score (defined later) based on the quality of their response. This allows the system to evaluate the model node’s behavior probabilistically without compromising efficiency or privacy.

\begin{figure}[t]
    \centering
    \includegraphics[width=0.96\linewidth]{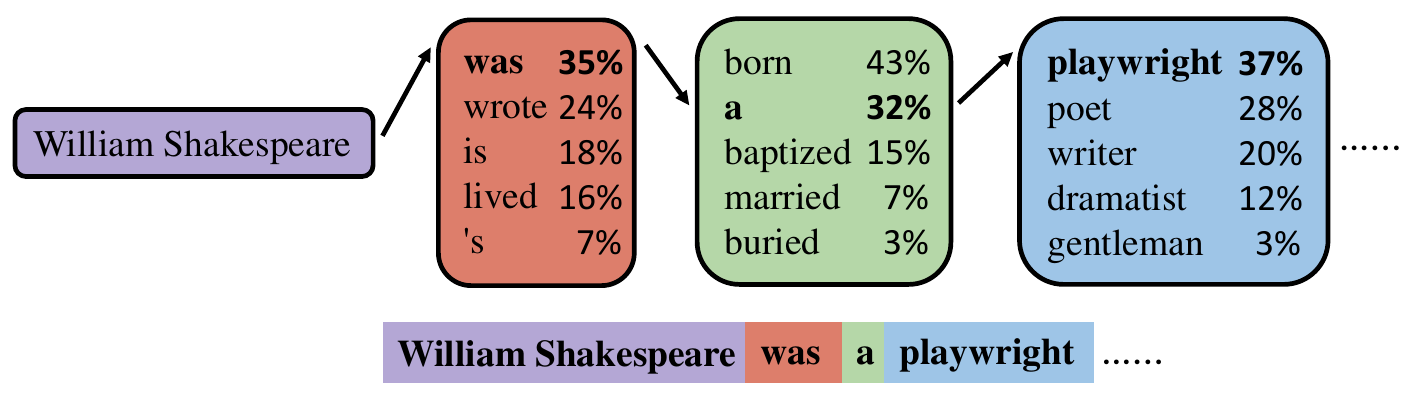}
    \vspace{-2ex}
    \caption{Verification for the model node challenge response.}
    \vspace{-3ex}
    \label{fig:verification}
\end{figure}

Each verification node locally deploys the same copy of the LLM as the model nodes. When receiving a challenge response from the model node, \red{verification nodes evaluate it against their local reference result, which is not necessarily the same.} The underlying principle is that, when using the same model and prompt, the conditional probability distribution over the next token is well-defined~\cite{fan2018hierarchical}. 
Suppose a verification node evaluates the correctness of a model node's response $r$ containing $n$ tokens, $r = (t_1, t_2, \ldots, t_n)$, to a challenge prompt. The verification node assesses the output token-by-token, as illustrated in Fig.~\ref{fig:verification}.
For each token $t_k$ at position $k$: 1) The verification node takes the prefix of the received response up to token $t_k$ and the original prompt. It uses its local LLM, conditioned on this input, to compute the probability distribution over all possible tokens at position $k+1$. 2) It then looks up the probability assigned by its local model to the actual token $t_{k+1}$ from the target model node. 
This process repeats for all tokens, generating a list of probabilities. 

The verification node calculates a score for each challenge response by computing its perplexity (PPL) based on the sequence of probabilities $PPL= \exp\Bigl(-\frac{1}{n}\sum_{i=1}^{n}\log p(t_i\mid t_{<i})\Bigr)$. The normalized perplexity $1/PPL$ is used for a single challenge.
By evaluating multiple challenges, the verification node obtains an average score $C(T)$ during time epoch $T$. 

The reputations score of a model node at time epoch $T$, $R(T)$, is updated by a \textit{moving average}: 
\vspace{-1.5ex}
$$R(T) = \alpha \cdot R(T-1) + \beta \cdot C(T)
\vspace{-1.5ex}$$
where $R(T-1)$ is the score from the last epoch, $C(T)$ is the average score from the challenges in $T$, and $\alpha$ and $\beta$ are the weights for the moving average. In our implementation, we set $\alpha=0.4$ and $\beta=0.6$. 

The punishment to the reputation for a low score should be \textit{much stronger} than the reward for a high score. This ensures the model node has to respond correctly consistently to earn a high reputation.   
We use a \textit{sliding window approach} to apply punishment to low values of $C(T)$. The verification nodes maintain a sliding window of each model including the past $W$ values of $C(T)$--the window size $W$ is set to 5 in our implementation. If $C(T)$ is smaller than a threshold $\tau$, it is considered an abnormal value. Let the count of abnormal values in the past $W$ be $c$. If $c/w$ is above a predefined threshold $\gamma$ (such as 1/5), the punishment applies as follows:
\vspace{-2ex}
$$R(T) = \alpha \cdot R(T-1) + \frac{W+1}{W+c/\gamma +2} \cdot C(T)
\vspace{-1.5ex}$$
Therefore, 
the reputation decreases excessively when detecting abnormal values. If a model node's reputation drops below a critical level (like 0.4), it will be marked as untrusted. 


Verification is performed epoch-by-epoch. In each epoch, the committee collaboratively assesses the correctness of model nodes and updates their reputation accordingly. 
PlanetServe relies on an open-source consensus protocol   Tendermint~\cite{buchman2018latest} to ensures safety and liveness under partial synchrony. 
%
Each update message should be signed by at least $2n/3+1$ nodes before commitment. 
In each epoch, a verification node is selected as the leader to send challenge requests to model nodes. 
To reduce overwhelming model nodes, only the leader of each epoch communicates with the model nodes, collects their responses, and broadcasts the results to the committee. Each verification node then locally verifies the results independently and votes on the updates.

The leader $L_i$ of the epoch $e_i$ is selected pseudo-randomly and verifiably towards the end of the previous epoch. 
Specifically, we use a Verifiable Random Function to select $L_i$ based on the final commit hash of epoch $e_{i-1}$.
At the end of epoch $e_{i-1}$, the committee also agrees on the set of model nodes to be verified in epoch $e_i$, $M_i$, and the corresponding challenge prompts for each of them. This is prepared by the leader $L_{i-1}$ in this epoch. The prompt is a unique, random natural text question, indistinguishable from normal user prompts. 
No two model nodes should be asked the same prompt to prevent collusion or replay attacks. 
This pre-determined model node and prompt list is to prevent the leader $L_i$ from behaving maliciously in epoch $e_i$ by selectively ignoring certain model nodes or assigning inconsistent prompts.

In epoch $e_i$, the leader $L_i$ initiates the verification process by sending the pre-determined challenge prompts to each model node in $M_i$. 
Each model node receiving a prompt generates a response output, which is a sequence of output tokens. The model node cannot tell if this is a prompt from clients or verification nodes, because all requests are delivered through the anonymous overlay. 
Once responses are received, $L_i$ computes a credibility score for each model node and generates a reputation score accordingly using the above method.
$L_i$ then broadcasts the signed challenge responses list and reputation scores to the committee for validation.

If the response and digital signature do not match or the model node does not respond, the leader can include an ``invalid response from model node $x$'' message in the broadcast message, but does not reduce $x$'s reputation. It is to prevent the leader from attacking a model node by dropping or altering the response. Only when more than $1/3$ of the committee reported invalid responses for the same model node, implying that  $x$ misbehaves on purpose, $x$'s reputation will be reduced.   

Upon receiving the information from the leader, each verification node performs an integrity check locally. It verifies whether the list of model nodes $M_i$ and associated prompts match the pre-determined prompt list, and checks the signatures and timestamps of each model node response. If any inconsistency is detected - e.g., different prompts or wrong signatures - 
verification for the current epoch is aborted. A new leader will be selected for the next epoch. 
For valid challenge responses, each verification node independently computes the credibility and reputation score using its local LLM, and compares it with the score proposed by the leader $L_i$. The reputation update follows Tendermint's two-phase voting protocol: Pre-Vote and Pre-Commit. 
If they match with negligible variance, the verification node sends a pre-vote. 



\begin{figure*}[t]
   \begin{minipage}[t]{0.33\textwidth}
        \centering
        \includegraphics[width=\linewidth]{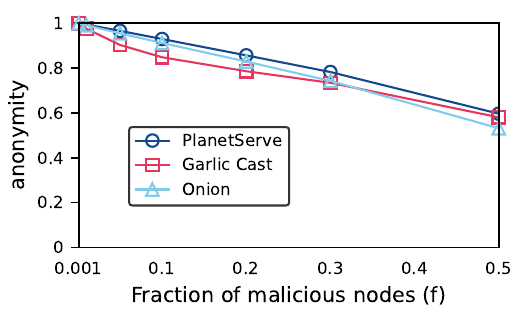}
        \vspace{-5ex}
        \caption{Anon. vs. malicious frac.}
        \vspace{-3ex}
        \label{fig:anonymity}
     \end{minipage}
    \begin{minipage}[t]{0.33\textwidth}
        \centering
        \includegraphics[width=\linewidth]{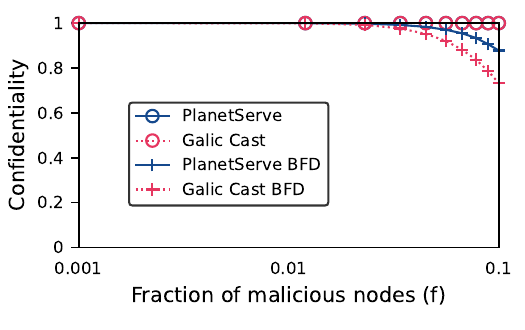}
        \vspace{-5ex}
        \caption{Conf. vs. malicious frac.}
        \vspace{-3ex}
        \label{fig:confidentiality}
    \end{minipage}
      \begin{minipage}[t]{0.33\textwidth}
        \centering
        \includegraphics[width=\linewidth]{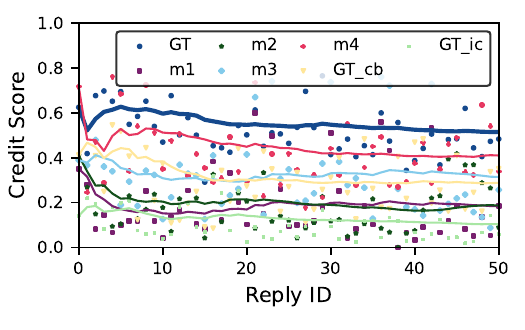}                
        \vspace{-5ex}
        \caption{\red{Credit score over models.}}
        \vspace{-3ex}
        \label{fig:cred}
     \end{minipage}
\end{figure*}

\vspace{-2ex}
\section{Security Analysis and Evaluation}
\label{sec:security}


\vspace{-2ex}
\subsection{Anonymity}
\vspace{-1ex}
We apply a commonly used entropy-based metric \cite{zhuang2005cashmere,garlic} to measure the anonymity of a system, whose definition is presented in the Appendix~\ref{sec:def:anony}. It ranges from 0 and 1, and a higher value represents better anonymity. 

We use simulations to compare the normalized entropy across three systems: onion routing \cite{Onion1998,Tor04}, garlic cast (GC) \cite{garlic}, and PlanetServe, in a 10,000-node network under different fractions of malicious nodes. As shown Fig.~\ref{fig:anonymity}, PlanetServe consistently achieves the highest level of anonymity, particularly in low-to-moderate corruption scenarios. For instance, at a corruption fraction of $f = 0.05$, the normalized entropy for PlanetServe reaches $0.965$, compared to $0.954$ for onion and $0.903$ for GC.
While introducing path diversity enhances resilience to individual path failures, it simultaneously increases the risk of de-anonymization if adversaries are capable of cross-path collusion. However, PlanetServe uses different path IDs for different paths, hence achieving greater robustness against colluding adversaries. 

\vspace{-2.5ex}
\subsection{Message confidentiality}
\label{sec:analysis:confidentiality}
\vspace{-1ex}
The content of a prompt or response is revealed to adversaries if the colluding adversaries exist on more than $k$ paths of the user node. where $k$ is the decoding threshold for S-IDA ~\cite{rivest1997all,stinson2001something, SIDA93}. 
Even if adversaries control $k$ paths, exhaustive brute-force decoding over all possible combinations of forwarded messages remains computationally prohibitive. 
In Fig.~\ref{fig:confidentiality}, both PlanetServe and GC show degradation in confidentiality if brute-force decoding is possible -- a big assumption on attack's capability --
with GC dropping to $0.73$ and PlanetServe to $0.88$ under high adversarial presence ($f=10\%$). 
If brute-force decoding is impossible, both systems maintain almost perfect confidentiality across all scenarios. 


\vspace{-2ex}
\subsection{Detecting Dishonest Model Nodes}
\label{sec:detect}
\vspace{-1ex}
We focus on detecting dishonest model nodes that use low-quality models or forged responses. %
Note that a model node cannot differentiate the requests from normal users and those from verification nodes, because all requests are routed through the anonymous overlay. 


We use \texttt{Meta-Llama-3.1-8B-Instruct-Q4\_0} as the ground truth (GT) model for evaluation, representing a \red{reference response}. To simulate dishonest or degraded behavior, we introduce four alternative models with lower capabilities or altered configurations. Specifically, \textbf{m1} refers to \texttt{Llama-3.2-3B-Instruct-Q4\_K\_M}, \textbf{m2} refers to \texttt{Llama-3.2-1B-Instruct-Q4\_K\_M}, \textbf{m3} refers to \texttt{Llama-3.2-1B-Instruct-Q4\_K\_S}, and \textbf{m4} refers to \texttt{Llama-3.2-3B-Instruct-Q4\_K\_S}. These models exhibit varying levels of performance degradation due to either reduced model size or quantization settings, and serve to evaluate the sensitivity and discriminative power of the PlanetServe verification and reputation framework. In addition, we include two further settings based on the ground-truth model but with deliberate prompt alterations. \textbf{gt\_cb}, where prompts are rewritten into clickbait-style headlines, and \textbf{gt\_ic}, where prompts are followed by injected long-form continuations spanning diverse genres (e.g., fictional expansions, documentary-like prose, or pop-psychology explanations). 

\begin{figure*}[t]
    \centering
    \begin{subfigure}[t]{0.32\textwidth}
        \centering
        \includegraphics[width=\linewidth]
        {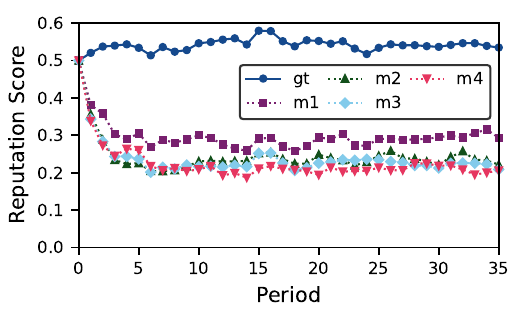}
        \vspace{-5ex}
        \caption{\red{Punishment threshold $\gamma = 1$.}}
        \vspace{-2ex}
        \label{fig:punish_level_1}
    \end{subfigure}
    \hfill
    \begin{subfigure}[t]{0.32\textwidth}
        \centering
        \includegraphics[width=\linewidth]        {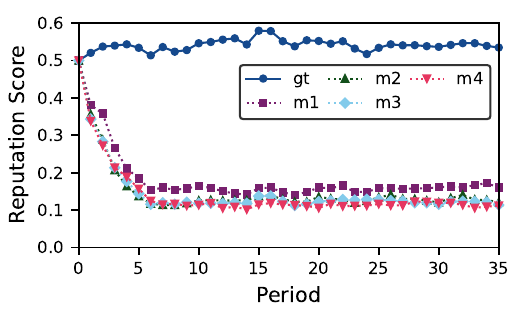}
        \vspace{-5ex}
        \caption{\red{Punishment threshold $\gamma = \frac{1}{3}$.}}
                \vspace{-2ex}
        \label{fig:punish_level_3}
    \end{subfigure}
    \hfill
    \begin{subfigure}[t]{0.32\textwidth}
        \centering
        \includegraphics[width=\linewidth]
        {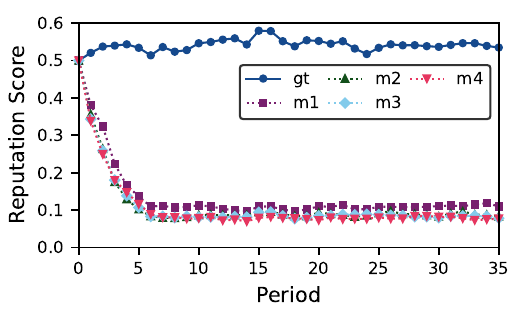}
        \vspace{-5ex}
        \caption{\red{Punishment threshold $\gamma = \frac{1}{5}$.}}
        \vspace{-2ex}
        \label{fig:punish_level_5}
    \end{subfigure}
    \caption{\red{Reputation changes with different punishment levels.}}
            \vspace{-1ex}
    \label{fig:punish_level}
\end{figure*}

We first run 50 prompts and evaluate the responses from all five models by measuring the credit scores in the normalized perplexity. 
Fig.~\ref{fig:cred} shows normalized perplexity-based credibility scores. Each point corresponds to the score of a single reply. We find that statistically, the ground truth model (GT) provides higher scores than other models. A false positive is possible, but in the long term, it will be corrected if the node follows the protocol and its credit score converges to the correct value.
\red{Fig.~\ref{fig:punish_level} illustrates how the reputation scores evolve over time across 35 epochs (50 prompts each) by varying three penalty sensitivity levels: $\gamma = 1$, $\frac{1}{3}$, and $\frac{1}{5}$. 
The results show a clear separation between GT and the weaker models after the first epoch, validating the effectiveness of the token-level probabilistic verification strategy.}
Fig.~\ref{fig:punish_level_1} shows the reputation trajectories under Level-1, where the penalty is relatively lenient. In this setting, the dishonest models experience slower degradation and tend to stabilize around reputation scores of 0.2 - 0.4, which may allow them to remain in the trusted pool for longer periods. With a stricter penalty in $\gamma = \frac{1}{3}$ Fig.~\ref{fig:punish_level_3}, reputation scores of dishonest models exhibit a more significant decline, often dropping below 0.2 by period 5. For the most aggressive punishment threshold $\gamma = \frac{1}{5}$ in Fig.~\ref{fig:punish_level_5}, results in an early and sharp decline of reputation scores, with most dishonest models falling below 0.1 within the first 5 periods. \red{Although the paper only shows a few set of experiments, we have run many similar experiments and PlanetServe can detect weaker models in every set of experiments.}

\begin{figure*}[t] 
    \begin{minipage}[t]{0.67\textwidth}
        \centering
    \begin{subfigure}[t]{0.477\linewidth}
        \centering
        \includegraphics[width=\linewidth]{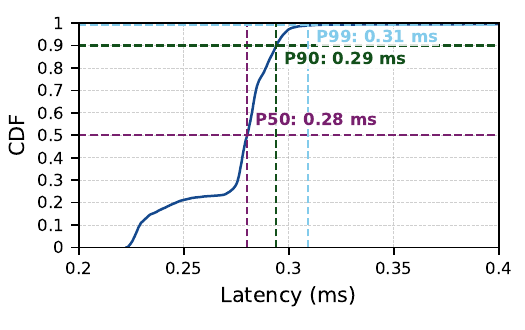}
        \vspace{-5ex}
        \caption{Preparation latency}
        \label{fig:prep_latency}
    \end{subfigure}
    \hspace{0.01\textwidth}
    \begin{subfigure}[t]{0.477\linewidth}
        \centering
        \includegraphics[width=\linewidth]{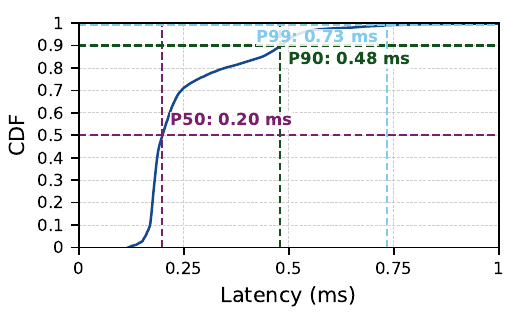}
        \vspace{-5ex}
        \caption{Decryption latency}
        \label{fig:decrypt_latency}
    \end{subfigure}
    \vspace{-2ex}
    \caption{Latency of preparing cloves on model nodes and decrypting cloves\\  on user nodes.}
    \vspace{-3ex}
    \label{fig:clove_latency}
    \end{minipage}
    \hfill
      \begin{minipage}[t]{0.32\textwidth}
        \centering
         \includegraphics[width=\linewidth]{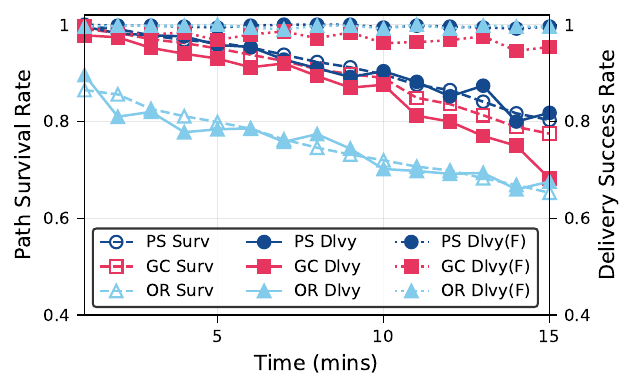}
         \vspace{-4ex}
        \caption{\red{Survival probability and delivery rate under churn (200 nodes/min).}}
        \vspace{-5ex}
        \label{fig:churn}
     \end{minipage}
\end{figure*}

Based on the empirical results, we set the reputation threshold to detect a dishonest model as 0.4 and $\gamma = \frac{1}{5}$. 

\vspace{-3.5ex}
\subsection{Other Security Properties}
\label{sec:othersec}
\vspace{-1ex}

\textbf{Denial-of-Service (DoS) attacks.} We consider three DoS scenarios. 
1) 
If the leader is malicious and refuses to send challenge prompts to model nodes or broadcasts challenge responses to the committee, the system simply fails the current epoch and restores the verification process in the next epoch with a new leader. 
2) If some committee members refuse to verify and vote, the system can still rely on the remaining honest verification nodes to vote and reach a BFT consensus.
3) A model node does not respond to user requests. This behavior can be detected by the verification protocol.  

\textbf{Counterfeiting.} There are three main types of counterfeiting attacks by a malicious verification leader.
1) The leader attempts to degrade some model nodes by sending them modified challenge prompts different from the agreed prompt list. 
Since model node responses always include the original prompt, other verification nodes can detect this deviation. 
2) The leader alters the responses from the model nodes before broadcasting them to the committee. In PlanetServe, each response from a model node includes its digital signature, which ensures the integrity and authenticity of the responses. Any tampering by the malicious leader would be easily detected. 
3) The leader can intentionally skip broadcasting responses from certain model nodes or falsely claim that a model node returned an ``invalid response". To prevent this, all verification nodes reach consensus in advance on the list of model nodes to be challenged for the epoch. If verification nodes detect that some expected responses are missing or marked as invalid without justification, they will initiate their own challenge prompts to those model nodes, which are distinct from the original prompts used by the leader, to prevent auditing detection by the model nodes. If more than 2/3 of the verification nodes receive valid responses from the model nodes, the system will identify the leader as malicious. 

\red{\textbf{Sybil attacks.} }
An adversary could potentially leverage Sybil attacks to control multiple verification nodes to alter verification results. 
\rv{PlanetServe requires verification nodes to pass credential checks and jointly validate committee decisions via BFT consensus, eliminating trust on a single authority. Verification nodes are operated by entities with real-world identities (e.g., recognized academic or research institutions), which imposes non-trivial cost on large-scale Sybil creation. Compromised nodes can be revoked and removed upon detection. Moreover, verification nodes are selected to be geographically and administratively diverse, making it difficult for an adversary to compromise $\geq 1/3$ of the committee. To further limit prolonged adversarial influence, committee members are periodically rotated through randomized re-selection, and misbehaving nodes are excluded.}




\vspace{-3ex}
\section{Performance Evaluation}
\label{sec:evaluation}

\vspace{-2.5ex}
\subsection{Methodology}
\label{sec:eval:method}
\vspace{-1ex}

\begin{figure*}[t]
    \centering
    \includegraphics[width=6.6in]{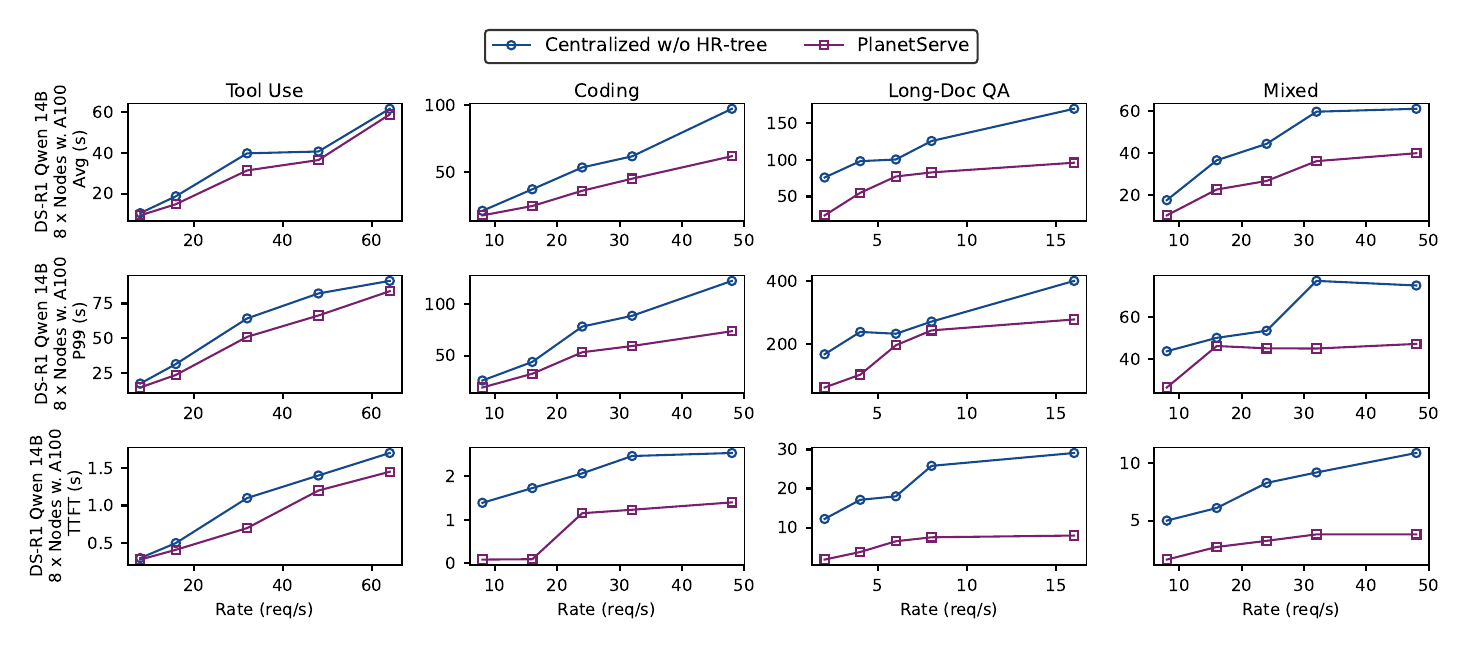}
    \vspace{-4ex}
    \caption{\red{Latency of LLM serving w/ and w/o HR-Tree.}} 
   \vspace{-3ex}
    \label{fig:dm}
\end{figure*}

We deploy the testbed of PlanetServe in a public cloud. The testbed includes eight model nodes with mid-tier hardware, each carrying one NVIDIA RTX A6000 GPU (48 GB) and serving a Meta-Llama-3 8B model, eight other model nodes with high-performance hardware, each carrying  one NVIDIA
A100 GPU (80 GB) and serving a \red{DeepSeek-R1-Qwen-14B} model. We run two machines as the verification nodes. One carries  an NVIDIA
A100 (40 GB SXM4), 30 virtual CPUs, 200 GiB of RAM, and
a 0.5 TiB SSD, and the other carries an NVIDIA GH200
(96 GB HBM), 64 virtual CPUs, 432 GiB of RAM, and 4 TiB
SSD.
There is no centralized coordinator. 
Internet latency in simulations is drawn from our measurements.
We use the latest version of vLLM \cite{vllm,vllmgithub}, a recently developed LLM serving library, on each single model node.  

Each user request is encoded into 4 cloves routed through 3-hop paths, with successful recovery requiring at least 3 cloves ($k=3$). 8-bit hash values are used for the HR-tree to optimize the balance between memory footprint and false-positive rate. The chunk-length arrays $L$ are refreshed every 10,000 requests, and HR-tree state synchronization occurs every 5 seconds. 
Queries are dispatched according to a Poisson distribution with varied mean inter-arrival times, accurately simulating real-world user query patterns and request bursts commonly observed in production environments.

The evaluation includes four realistic workloads:

1. \textbf{ToolUse (ToolBench\cite{guo2024stabletoolbench})}: 
210k queries with tool-specific instructions; prompts average 7,206 tokens after Zipf-1.1 sampling. Prefix sharing is moderate, and outputs are capped at 100 tokens per query for the models we evaluate.

2. \textbf{Coding (APPS\cite{hendrycksapps2021})}: 
10,000 coding problems with detailed solution requests; prompts average 1,802 tokens after Zipf-0.8 sampling. Prefix overlap is minimal, and outputs are capped at 1,000 tokens per query for the models we evaluate.

3. \textbf{Long-Doc QA (LooGLE\cite{li2023loogle})}: 
776 long documents paired with 6.4k questions. Each prompt includes a long document as a prefix followed by a question, averaging 10,985 tokens per prompt after Zipf-0.6 sampling. Outputs are capped at 100 tokens per query for the models we evaluate.

4. \textbf{Mixed Workload}: 
It integrates the ToolUse, Coding, and Long-Doc QA workloads in a 3:6:1 ratio per real-world traces~\cite{Preble25,wu2023fast}, averaging 9,959 tokens per prompts (All token counts above are computed using the Llama 3 tokenizer).


\vspace{-3ex}
\subsection{Anonymous Routing}
\label{sec:eval:routing}
\vspace{-1ex}
\begin{figure*}[tbp]
    \centering
    \begin{minipage}[t]{0.33\linewidth}
        \centering
        \includegraphics[width=\linewidth]{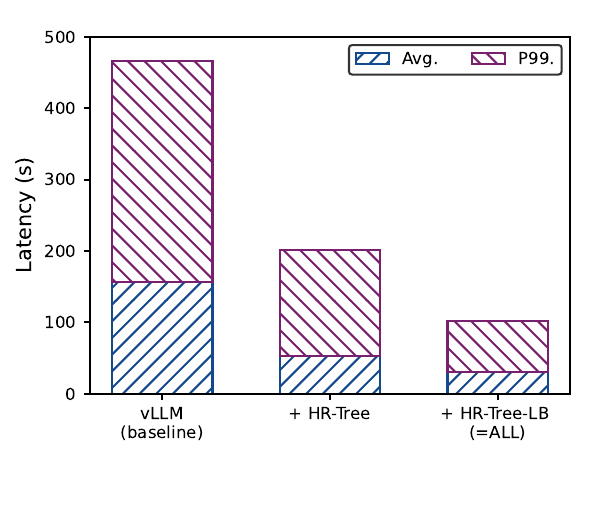}
    \vspace{-9ex}
    \caption{\red{Ablation study of latency.}} 
    \label{fig:ablation}
    \end{minipage}
    \vspace{-2ex}
    \hfill
    \begin{minipage}[t]{0.33\linewidth}
        \centering
        \includegraphics[width=\linewidth]{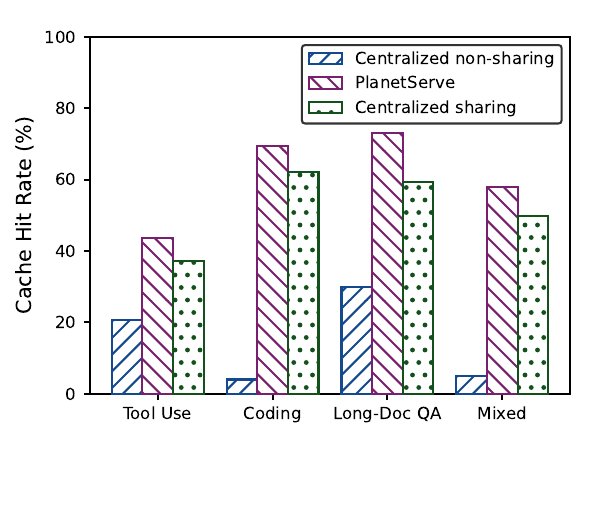}
        \vspace{-9ex}
        \caption{\red{Cache hit rate.}}
        \label{fig:cache_hit_rate}
    \end{minipage}
    \vspace{-2ex}
    \hfill
    \begin{minipage}[t]{0.33\linewidth}
        \centering
        \includegraphics[width=\linewidth]{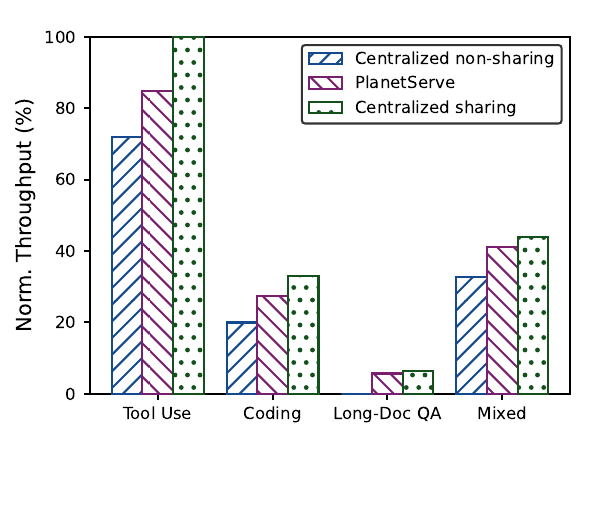}
        \vspace{-9ex}
        \caption{\red{Norm. LLM throughput}}
        \label{fig:normalized_throughput}
    \end{minipage}
    \vspace{-3ex}

\end{figure*} 

We evaluate the performance overhead of anonymous message (clove) processing using the Toolbench dataset~\cite{guo2024stabletoolbench}. Clove preparation is performed on a model node equipped with a single NVIDIA A100 40GB SXM4 GPU, 30 vCPUs, 200~GiB RAM, and 0.5~TiB SSD running Llama-3.3-70B. 
Decryption is executed on a user node with a single Intel i7-7700 CPU @ 3.60GHz. Figure~\ref{fig:clove_latency} shows the CDFs for both preparation and decryption latency over 10{,}000 trials.

Clove preparation is lightweight, with mean and P50 latencies of 0.273 and 0.280~ms. Even at the P99 percentile, latency remains below 0.31~ms, indicating tight bounds. Clove decryption is similarly efficient, with a slightly higher mean latency of 0.302~ms and a P50 of 0.309~ms. The P99 latency is 0.334~ms, while maintaining a 100\% success rate.
Overall, both operations exhibit low and stable latency.

\red{We simulate communication survival under churn in a 3,119-node network and a churn rate of 200 nodes/min -- a very high rate and
 deployed 14 AWS t3.micro instances across multiple regions and measured end-to-end routing latency with realistic prompt payloads over 4,000 runs (3 nodes randomly fail in the USA; 1 randomly fails worldwide). 
The simulation incorporates latency, link failures, packet loss, and congestion (Fig.~\ref{fig:churn}). PlanetServe achieves the highest path survival rate, maintaining high delivery under failures, while Onion degrades significantly.
}


\vspace{-3ex}
\subsection{\red{Latency of Confidential Computing (CC)}}
\vspace{-1ex}
\label{sec:CC-latency}

\begin{table}[t]
\centering
\begin{tabular}{ccccc}
\hline
\textbf{model} & \multicolumn{2}{c}{\textbf{mean}} & \multicolumn{2}{c}{\textbf{P99}} \\ \hline
 & CC-on & CC-off & CC-on & CC-off \\ \hline
Llama-3.1 8B & 132.19 & 130.95 & 123.94 & 122.70 \\ \hline
DS-R1-Q 14B & 211.58 & 210.96 & 197.9527 & 195.77 \\ \hline
\end{tabular}
\vspace{-1ex}
\caption{\red{Latency comparison under CC mode.}}
\vspace{-3.5ex}
\label{tab:cc_latency}
\end{table}

\red{Table~\ref{tab:cc_latency} reports results on 
Azure Standard NCC40ads H100 v5 VMs with CC enabled and the corresponding NC40ads H100 v5 VMs without CC, both equipped with a single NVIDIA H100 GPU, 40 vCPUs, and 320~GB RAM at 20 requests per second.
With the request rate fixed at 20 requests per second in both settings, CC introduces minimal latency overhead.
}

\vspace{-3ex}
\subsection{Overlay Forwarding}
\label{sec:forwardExp}
\vspace{-2ex}

We evaluate three types of latency for the model node responding to user requests: average generation latency capped at 1000 tokens (Avg), 99th percentile (P99), and Time to First Token (TTFT).
The LLMs used in the evaluations are Meta Llama 3 (8B) running on A6000 and DeepSeek-R1 (14B) running on A100. 
\red{We compare PlanetServe with a centralized baseline of 8 A100 GPUs without KV-cache sharing.}
Each data point reported here is the average of three independent runs of the entire workload. 
Latency results are shown in Fig.~\ref{fig:dm}  for PlanetServe and the \red{centralized baseline}. \rv{A detailed comparison against centralized serving with and without KV-cache sharing is provided in Fig.~\ref{fig:u4} in Appendix A.~\ref{sec:A100results}.}
PlanetServe consistently demonstrates significant latency improvements across all evaluated workloads, LLM, and GPU setups. 
Notably, under heavy workload (LooGLE and Mixed), the difference is more evident, indicating the effectiveness of PlanetServe’s overlay forwarding and cache-reuse strategy. The TTFT metric directly impacts perceived user responsiveness, a critical factor in practical deployments. 
PlanetServe provides significantly shorter latency in TTFT.  
For both LLMs, PlanetServe reduces TTFT at higher request rates by 40-50\%. 
In addition, for higher request rates, PlanetServe maintains slower latency growth. 

To isolate design benefits, we perform an ablation study by incrementally enabling HR-tree and load balancing on 8 A100 GPUs as model nodes running Llama-3.1-8B. Figure~\ref{fig:ablation} shows results under the ToolUse workload with a Zipf-1.1 distribution. HR-tree reduces both average and P99 latency by over 50\%, with load balancing providing further gains.

We show the KV cache hit rates of the mixed workload on \red{DeepSeek-R1-Qwen-14B} model nodes in Fig.~\ref{fig:cache_hit_rate}. 
\red{In addition to the centralized baseline, we further compare PlanetServe to centralized serving with 8× A100 GPUs, running vLLM with the default continuous-batching scheduler and tensor parallelism.} 
PlanetServe consistently achieves significantly higher cache hit rates across all evaluated workloads. 
The throughput comparison is
shown in Fig.~\ref{fig:normalized_throughput}. 
\red{PlanetServe outperforms the baseline without sharing, and tensor parallelism provides the highest throughput due to the scheduler and parallelism.} 


\vspace{-2.5ex}
\subsection{Verification throughput}  
\vspace{-2ex}


Assuming each model node requires verification 50 times per day and a deployment ratio of up to 100 model nodes per VN, the required throughput is ~208 verifications per VN per hour.
We evaluate the verification nodes on two platforms. 
A high-end GH200 system achieves 45.04 verifications per minute, while a standard A100 system reaches 20.72 per minute. Both configurations meet the required verification throughput.

\vspace{-4ex}
\section{\red{Discussion}}
\label{sec:discuss}
\vspace{-2ex}

\red{\textbf{Lesson learned from past projects.} 
We investigate the challenges of past decentralized overlay networks, such as PlanetLab and P2P file sharing, and discuss how the design of PlanetServe can learn from these projects. 1) Incentive design is important. Performance efficiency alone does not necessarily lead to adoption. Hence, we design a comprehensive incentive infrastructure with verification services.  2) Churn resilience and security should be considered at the beginning of deployment, which are the main research objectives of this work. 3) Integration with modern computing architecture. Some past P2P networks cannot integrate with the cloud architecture, which significantly limits their applications. The design of PlanetServe aims at a flexible and scalable architecture compatible with cloud and can evolve with new technologies.} 

\red{\textbf{More complex models.} 
This paper focuses on text-only tasks. Future research can extend the architecture to support more complex models. 
One direction is multimodal support. HR-Tree naturally supports non-textual data by storing hash values rather than raw text, offering advantages over existing KV cache sharing methods.
Another is to support advanced reasoning and tool-using models, whose workloads may require adapting PlanetServe's current KV cache sharing and load-balancing mechanisms. 
The core PlanetServe infrastructure remains unchanged and can be incrementally optimized.}

\red{\textbf{Model partitioning.} 
Due to space constraints,
we do not consider model partitioning  \cite{borzunov2023petal}, which is complementary to PlanetServe. 
This combined direction is left for future work.} 

\red{\textbf{Blockchain integration.}
Our committee-based BFT mechanism offers efficient serving and verification without blockchain overhead.
Blockchain offers transparency, auditability, and global coordination via smart contracts, and can be integrated to support large-scale deployments.
}

\rv{\textbf{Practical deployment considerations.}}
\rv{We validate PlanetServe’s mechanisms through moderate-scale deployments and large-scale simulations capturing heterogeneous failures. The architecture scales without centralized control, though larger deployments may introduce additional trade-offs.}

\vspace{-4ex}
\section{\rv{Limitations}}
\label{sec:limitations}
\vspace{-2ex}
\rv{\textbf{Model integrity verification using trusted execution environments.}
PlanetServe uses confidential computing to protect prompt privacy, which is orthogonal to the model verification mechanism. Hardware-based attestation could strengthen model integrity by verifying that specific model weights are loaded within a trusted execution environment. Prior work such as AttestLLM~\cite{zhang2025attestllm} explores this direction for on-device, single-node LLM deployments, but does not address distributed, multi-node, or decentralized serving environments.}

\rv{\textbf{Functional alignment verification and model identity.}
PlanetServe’s verification prioritizes functional alignment, validating whether responses match the expected behavior and quality of the intended model rather than enforcing guarantees on model identity. Thus, it cannot distinguish between different models that produce statistically similar outputs. This limits use cases such as benchmarking that require attributing outputs to a specific implementation. Ensuring model identity without hardware support remains an open challenge.}
\vspace{-4ex}
\section{Related Works}
\label{sec:related}
\vspace{-2ex}
LLM serving engines and software frameworks 
have been studied and developed \cite{vllm,PowerInfer,sglang24,Speculative,Preble25,TensorRT,TGI,Fairness2024Sheng,LoongServe2024,DistServe2024,SplitWise2024,CacheGen2024,liu2024droidspeak,InfiniGen}. One key challenge of an LLM serving engine is the management of KV cache. For example, vLLM \cite{vllm} stores KV cache as non-continuous blocks, each with a fixed number of tokens. It enables sharing KV caches for different user requests with similar prompts.      

Distributed LLM serving studies the management of a group of serving nodes in a cloud/cluster environment \cite{ServerlessLLM,IntelligentRouter,sglang24,Preble25,CacheGen2024}. One key research problem is to distribute requests to different serving nodes \cite{IntelligentRouter,sglang24,Preble25}. 
SGLang \cite{sglang24} and Preble \cite{Preble25} use a radix tree to find reusable KV caches across different user requests and implement scheduling algorithms to achieve load balancing. DistServe \cite{DistServe2024} assigns prefill and decoding to different nodes to eliminate their interferences. These studies assume a centralized ``scheduler'' or ``router'' that analyzes all user requests based on various scheduling algorithms and sends the requests to serving nodes. In contrast, there is no such central scheduler in PlanetServe with knowledge of the entire system. PlanetServe is the first study of the three key problems in decentralized serving. 

Petals \cite{borzunov2023petal} proposes to run an LLM (such as BLOOM-176B) on multiple consumer GPUs collaboratively. The Knowledge Delivery Network (KDN) \cite{kdn} allows storing and exchanging KV caches among different serving engines. \red{DiLoCo \cite{diloco2024} uses distributed clusters for training. Orca \cite{Orca2022} uses a centralized scheduling algorithm to improve serving latency and throughput. None of them address the challenges discused in PlanetServe, including decentralized serving and verification, overlay forwarding, and anonymous communication.}


\vspace{-4ex}
\section{Conclusion}
\label{sec:conclusion}
\vspace{-2ex}
This work introduces PlanetServe, a novel decentralized infrastructure for scalable and privacy-preserving LLM serving. We present designs addressing four key challenges: overlay network organization, anonymous communication, overlay forwarding for efficiency, and verification of model serving quality. These issues are fundamental and inherent to future decentralized LLM-serving system. We implement a prototype of PlanetServe and evaluation results show that PlanetServe achieves low serving latency with minimal overhead for security and trustworthy features. 
We believe this work offers a new direction that will attract further research and eventually help democratize future AI innovations and deployment. 
\vspace{-4ex}
\section{Acknowledgement}
\label{sec:acknowledgement}
\vspace{-2ex}
We thank our shepherd Walter Willinger and the anonymous reviewers for their suggestions and comments.
The authors were partially supported by National Science Foundation Grants 2322919, 2420632, 2426031, and 2426940.


\bibliographystyle{plain}
\bibliography{reference}
\section*{Artifact Appendix}
We release an open-source prototype of PlanetServe,
together with the demos and evaluation scripts.
The source code is publicly available at:
\url{https://github.com/fffeifang/PlanetServe.git}.

The repository includes a top-level README that provides an overview of the PlanetServe system, repository structure,
and evaluation roadmap. Each major component is further documented with a dedicated README describing its purpose,
usage, and experimental setup of the corresponding component.
\section*{Appendix}

\subsection*{A1. Limitations of existing overlay anonymous routing.}
\label{sec:limit-onion}
\textbf{Limitations of overlay Onion routing.} 
Onion routing \cite{Onion1998} was proposed to achieve anonymous routing in overlay networks, such as Tor \cite{Tor04}. If we apply the same idea to the overlay network in PlanetServe, \red{there are a few main challenges. First, the first layer of Onion nodes (called as guards in Tor) will definitely know the sending user of a request, which should be avoided in PlanetServe. Second,} the failure probability of an Onion path increases exponentially with the path length, especially for an overlay network of users with a high churn rate. In addition, each relay needs to perform public-key decryption on the entire message, introducing non-trivial overhead for users. 

\textbf{Strengths and weaknesses of slicing protocols.} One idea is to 
use sliced messages for overlay anonymity \cite{IS-NSDI07,garlic}. In garlic routing \cite{garlic},
the sender sends $n$ slices, each is about $1/k$ of the original message size long and is called a clove. If the receiver gets $k$ out of $n$ cloves, it can recover the original message. Garlic achieves 1) anonymity without encryption and decryption operations on the paths; 2) reliability under node failures; 3) the bandwidth cost does not increase significantly with the $n-k$ redundant cloves.  
Garlic routing has a few weaknesses if we apply it directly to PlanetServe.
1) Garlic relies on random walks to establish routing paths and the corresponding ``proxies''. However, random walks add uncertainty to the success of finding enough proxies. 2) Garlic preserves the provider (model node) anonymity, which is unnecessary for PlanetServe.

\begin{algorithm}
\caption{HashRadix Tree Search}
\label{alg:hashradix}
\begin{algorithmic}[1]
\Function{Search}{HRTree, request, $\tau_c$}
        \State $d \gets 0$
        \State $l_1, l_2, ..., l_n \gets HRTree.L$
        \State $k \gets 0$
        \For{$i \gets 1$ to $n$}
            \State $chunk_i \gets hash(request[k..k+l_i], HRtree.mod)$
            \State $k = k + l_i$
        \EndFor
        \State $node \gets HRTree.root$
        \For{$i \gets 1$ to $n$}
            \If{$node$ has child $chunk_i$}
                \State $node \gets node.children[chunk_i]$
                \State $d \gets d + 1$
            \Else
                \State \Return $node.modelNodeList, d$
            \EndIf
        \EndFor
        \If{$d < \tau_c$}
        \State \Return $\emptyset$
        \Else
        \State \Return $node.modelNodeList, d$
        \EndIf
\EndFunction
\end{algorithmic}
\end{algorithm}
\subsection*{A2. Pseudocode}

Algorithm  \ref{alg:hashradix} presents  the pseudocode of the HR-Tree search algorithm.

\begin{algorithm*}
\caption{Load Balancing}
\label{alg:loadbalancing}
\begin{algorithmic}[1]
\Function{LoadBalancing}{HRTree, request, $\tau_c$} 
    \State $modelNodeList \gets search(HRTree, request, \tau_c)$
    \If{$modelNodeList = \emptyset$} \Comment{Cache miss}
    \State Forward to model node with fewest relative requests
    \Else \Comment{Cache hit}
    \State $candidate \gets$ Model node with fewest relative requests in $modelNodeList$
    \If{$candidate.load < candidate.threshold$}
    \State Forward to $candidate$
    \Else 
    \State Forward to one with fewest relative requests in all model nodes \Comment{Fallback to load balancing}
    \EndIf
    \EndIf
\EndFunction
\end{algorithmic}
\end{algorithm*}

Algorithm \ref{alg:loadbalancing} presents  the pseudocode of the load-balancing algorithm with the HR-tree result. 

\begin{algorithm*}
\caption{Model Verification}
\label{alg:verify}
\begin{algorithmic}[2]
\Function{CheckCredibility}{prompt, output} \Comment{Tokenized}
        \State $probabilities \gets []$
        \State $context \gets prompt$
        \For{$i \gets 0$ to $length(output)-1$}
            \State $next\_token \gets output[i]$
            \State $logprobs \gets \text{GetCompletionLogprobs}(context, max\_tokens=1)$
            \State $token\_prob \gets \text{FindMatchingProbability}(logprobs, next\_token)$
            \If{$token\_prob$ exists}
                \State $probabilities.append(token\_prob)$
            \Else
                \State $probabilities.append(\epsilon)$ \Comment\{Small constant\}
            \EndIf
            \State $context \gets context + next\_token$
        \EndFor
        \State \Return $score \gets \text{NormalizedPPL}(probabilities)$
\EndFunction
\end{algorithmic}
\end{algorithm*}

Algorithm \ref{alg:verify} presents the pseudocode of the model verification process. 

\begin{figure}[t]
    \centering
    \includegraphics[width=0.98\linewidth]{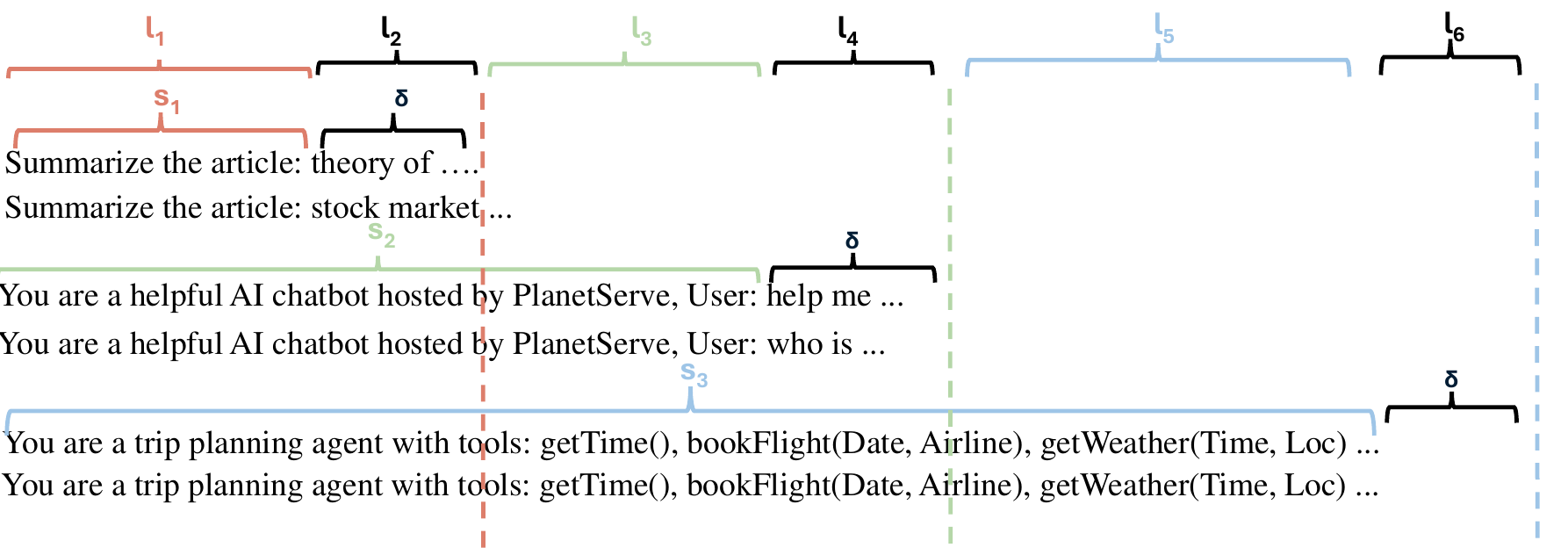}
    \vspace{-1ex}
    \caption{Chunk length array $L$ set by Sentries.}
    \vspace{-2ex}
    \label{fig:sentry}
\end{figure}

\subsection*{A3. The Sentry algorithm for the chunk length array}
Determining the optimal configuration for the chunk length array $L$ presents a significant challenge. Setting the chunk sizes too small results in a HashRadix Tree (HR-Tree) with excessive depth, potentially over-differentiating prefixes. Conversely, if chunk sizes are too large, requests with substantial prefix overlap and high cache affinity might be incorrectly treated as entirely distinct. To address this, PlanetServe incorporates a Sentry module. This component is responsible for collecting incoming requests and detecting common patterns, such as system prompts, which can inform the dynamic adjustment or initial configuration of the chunking strategy defined by $L$. How to set $L$ to effciently utlize the HR-Tree? As described in Fig. \ref{fig:sentry}, we define $S = s_1, s_2, \dots, s_n$ be the lengths of the distinct common system prompts detected by the Sentries, sorted in ascending order. The Sentry module uses these lengths, along with a small, fixed separator length $\delta$, to construct the chunk length array L used in Algorithm 1. The array L is defined as follows:
\begin{align}
    l_1 &= s_1 \\
    l_{2n} &= \delta \\
    l_{2n + 1} &= s_n - s_{n - 1} - \delta
\end{align}

This structure ensures that each distinct system prompt corresponds to a specific path segment in the HR-Tree, separated by a small $\delta$ chunk. This allows the initial layers of the HR-Tree to be efficiently utilized for routing based on common prompts, thereby enhancing cache affinity for requests that share these initial structures. Subsequent parts of the request (beyond the detected prompts) are tokenized using any remaining lengths in $L$ or a default chunking strategy.

\subsection*{\red{A4. Analysis of choice of $n$ and $k$ in anonymous routing.}}
\red{In anonymous routing of PlanetServe, we need $k$ of $n$ cloves to arrive for a successful communication. There are $l=3$ relays on each path. Assume the failure rate of an overlay node during the time of a round of communication is $f$. The probably that one path succeeds is $(1-f)^3$.
The rate that at least $k$ out $n$ succeed is:
$$P(X \ge k) = \sum_{i=k}^{n} \binom{n}{i} (1-f)^{3i} (1-(1-f)^3)^{n-i}$$}

\red{Using $n=4$ and $k=3$, even with a failure rate as high as $3\%$, the success rate is $>95\%$.}

\subsection*{A5. Definition of Anonymity Metric}
\label{sec:def:anony}
\textbf{Definition.} Let $S$ be the set of all nodes in the network with $|S| = N$. An attacker assigns to each node $x \in S$ a probability $p_x$ of being the source (or destination) of a message. The entropy of the system is defined as:
\[
H(S) = - \sum_{x \in S} p_x \log_2(p_x).
\]
Perfect anonymity occurs when the attacker has no information about the system, and thus must assign a uniform probability of $1/N$ to each node. In this case, the entropy reaches its maximum:
\[
H_{\max} = \log_2(N).
\]
To evaluate the anonymity of the system, we normalize the entropy as a metric. The normalized anonymity is given by:
\[
\frac{H(S)}{H_{\max}} = \frac{- \sum_{x \in S} p_x \log_2(p_x)}{\log_2(N)}.
\]
We use the above metric but need to know how to compute $p_x$. 
We assume a fraction $f$ of all users are collaborative malicious users. Some of them may sit on the $k$ paths between the client and the proxies. However, since they have no information on the prompt, they cannot tell if they are serving the same client. Suppose the number of all nodes on the $k$ paths is $L$. There may exist multiple chains of consecutive attackers on the paths. The attackers may guess that the predecessor of every chain is the source and the successor of every chain is the destination. Hence the probability of a guess being correct is $\frac{1}{L+1-f\cdot L}$. Let $\Gamma$ be the set of predecessors of all malicious chains. We then have,
\[
\Pr(x = \text{src}) = 
\begin{cases}
\frac{1}{L+1-f\cdot L} & \text{if } x \in \Gamma \\
(1-\frac{|\Gamma|}{L+1-f\cdot L}) \frac{1}{(1-f)N-|\Gamma|} & \text{otherwise} 
\end{cases}
\]
The entropy-based anonymity can be computed using the above probabilities.

\begin{figure}[t]
    \centering
    \includegraphics[]{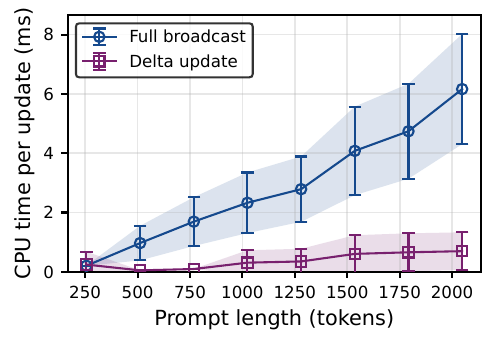}
    \vspace{-5ex}
    \caption{Computation cost for HR-tree update} 
   \vspace{-3ex}
    \label{fig:u1}
\end{figure}

\begin{figure}[t]
    \centering
    \includegraphics[]{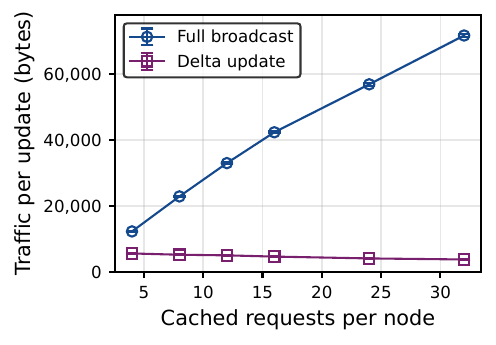}
    \vspace{-5ex}
    \caption{Network cost for HR-tree update} 
   \vspace{-3ex}
    \label{fig:u2}
\end{figure}

\begin{figure*}[t]
    \centering
    \includegraphics[width=6.6in]{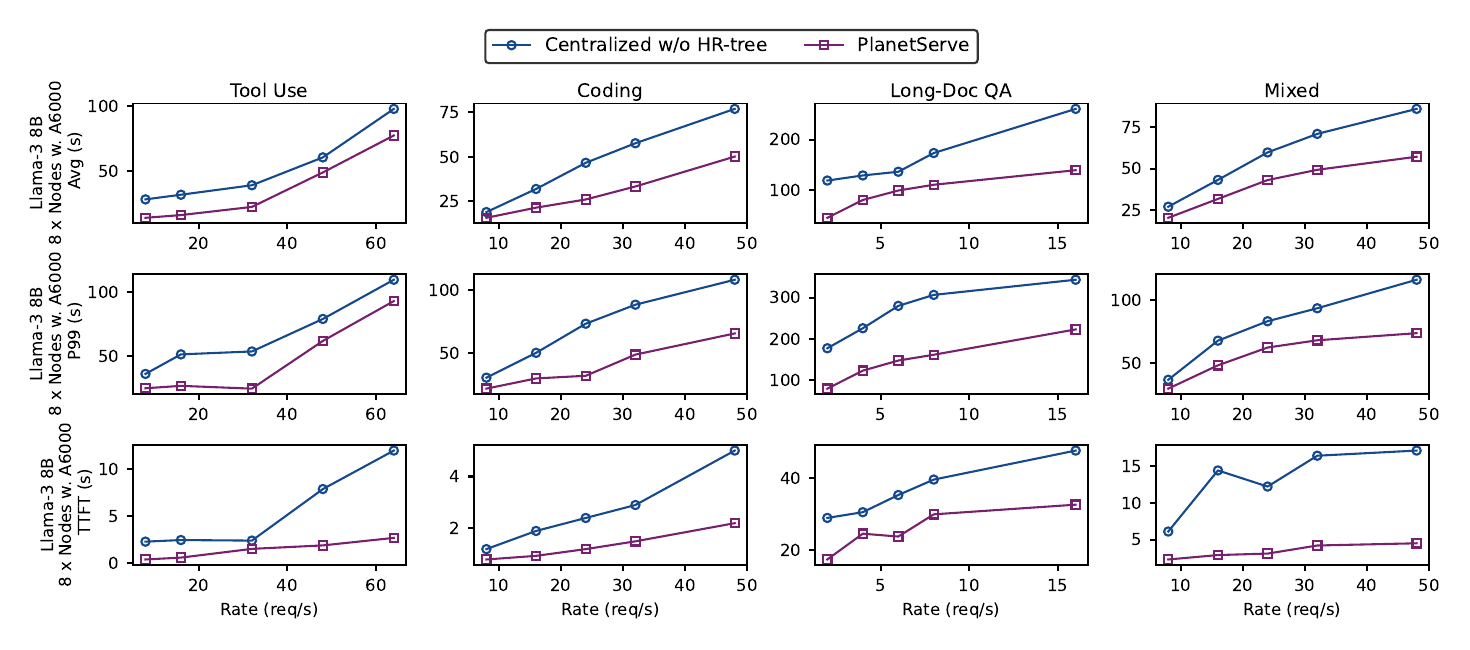}
    \vspace{-2ex}
    \caption{Latency of LLM serving with A100 GPUs} 
   \vspace{-3ex}
    \label{fig:u3}
\end{figure*}

\subsection*{\red{A6. HR-tree Update Cost}}
\label{sec:updatecost}
\red{We evaluate the computation cost and network cost of HR-tree update by comparing the full broadcast design and the proposed delta update and show the results in Fig.~\ref{fig:u1} and Fig.~\ref{fig:u2}, respectively. The results show that delta update significantly reduces the cost. }

\subsection*{\red{A7. Latency of LLM serving with A100 GPUs}}
\label{sec:A100results}
\red{We evaluate the latency of LLM serving with A100 GPUs and show the results in Fig.~\ref{fig:u3}. Compared to the A100 results in Fig.~\ref{fig:dm}, PlanetServe shows similar advantages. }

\begin{figure*}[t]
    \centering
    \includegraphics[width=6.6in]{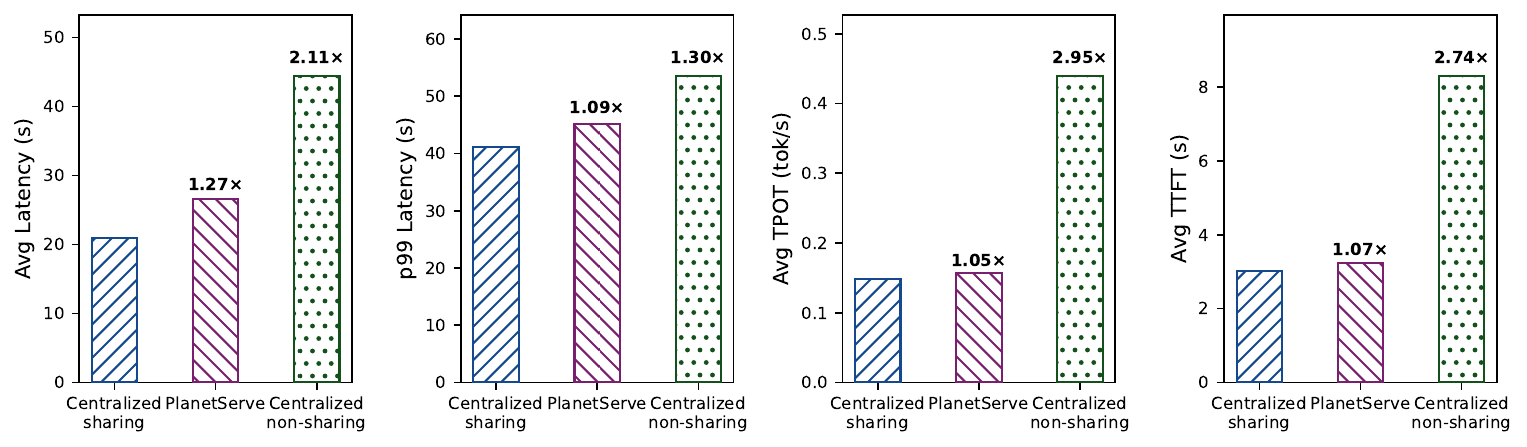}
    \vspace{-2ex}
    \caption{mixed with upper-bound} 
   \vspace{-3ex}
    \label{fig:u4}
\end{figure*}

\subsection*{\red{A8. Latency Comparison with Centralized Baselines}}
\label{sec:A100results}
\red{We compare PlanetServe with a centralized baseline of 8 A100 GPUs without KV-cache sharing and  centralized serving with 8 A100 GPUs, running vLLM with the default continuous-batching scheduler and tensor parallelism. Fig.~\ref{fig:u4} shows the latency comparison. We find that PlanetServe's latency is close to the Centralized sharing method, but significantly shorter than that of Centralized non-sharing. }

\subsection*{\red{A9. Resiliency to Intersection Attacks}}
\red{An intersection attack \cite{Intersection-CCS13} can be launched to anonymity systems if the target uses pseudonymity and the attacker has access to the user's history of visits and ISP logs. PlanetServe treats each sequence of prompts as independent. Hence, users do not use pseudonymity and the attacker cannot obtain user's history. Even though the attacker can access ISP logs (very big assumption!), they cannot match the user behavior among the massive amounts of Internet users. }

\subsection*{\red{A10. Real-World Routing Latency Measurements}}

\begin{figure}[t]
    \centering
    \includegraphics[width=0.96\linewidth]{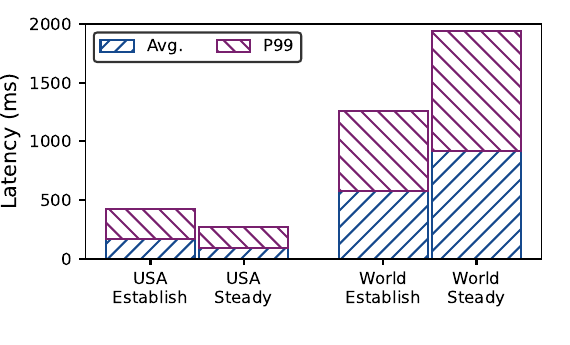}
    \vspace{-2ex}
    \caption{Measured Session-Establish and In-Session Latency Across Regions}
    \vspace{-3ex}
    \label{fig:latency_update}
\end{figure}

\red{To evaluate the practical impact of network overhead, we measured routing latency on real cloud infrastructure. Specifically, we deployed 14 AWS \texttt{t3.micro} instances, with each hop placed in a different region (datacenter), and used a mixed workload to simulate real prompt payloads. Over the course of 4000 runs, we observed three failures in the across-USA setting and one failure in the across-world setting.}

\red{Results are summarized in Fig.~\ref{fig:latency_update}. In the across-USA experiment (four regions in the United States), the session-establishment latency was 168.9 ms (P99: 256.8 ms), while the steady in-session latency was 92.9 ms (P99: 179.2 ms). In the across-world experiment (five regions across North America, Asia, Europe, and South America), the session-establishment latency was 577.4 ms (P99: 685.8 ms), and the steady in-session latency was 919.6 ms (P99: 1025.5 ms). }

\red{Even at inter-continental scales, these latencies are modest compared to LLM inference time (see Fig.~\ref{fig:dm}), suggesting that communication does not dominate end-to-end serving time. \rv{Nevertheless, this overhead reflects an inherent privacy--latency trade-off and may still be perceptible in latency-sensitive interactive settings.}}
\FloatBarrier
\clearpage 

\end{document}